\documentclass[aps, prd, superscriptaddress, preprintnumbers, floatfix, longbibliography, twocolumn, 10pt]{revtex4-2}

\usepackage{graphicx,amsfonts,amsmath,amssymb,amstext}
\usepackage{float,wrapfig}
\usepackage{subfigure, psfrag}
\usepackage{dsfont}
\usepackage{xcolor}
\usepackage[colorlinks=true,urlcolor=blue,citecolor=blue,linkcolor=blue]{hyperref}
\usepackage{bm}
\usepackage{mathtools}

\definecolor{purple}{rgb}{0.8,0,0.6}
\definecolor{darkgreen}{rgb}{0.00,0.6,0.00}
\newcommand{\revision}[1]{\textcolor{red}{#1}}

\extrafloats{100}

\begin{document}

\title{Electric and chiral response to a  pseudoelectric field in Weyl materials}
\date{July 23, 2022}

\author{A.~A.~Herasymchuk}
\email{arsengerasymchuk@gmail.com}
\affiliation{Department of Physics, Taras Shevchenko National Kyiv University, Kyiv, 03022, Ukraine}

\author{P.~O.~Sukhachov}
\email{pavlo.sukhachov@yale.edu}
\affiliation{Department of Physics, Yale University, New Haven, Connecticut 06520, USA}

\author{E. V. Gorbar}
\email{gorbar@bitp.kiev.ua}
\affiliation{Department of Physics, Taras Shevchenko National Kyiv University, Kyiv, 03022, Ukraine}
\affiliation{Bogolyubov Institute for Theoretical Physics, Kyiv, 03143, Ukraine}

\begin{abstract}
The electric and chiral current response to the time- and coordinate-dependent pseudoelectric field $\mathbf{E}_5$ in Weyl semimetals is studied. It is found that $\mathbf{E}_5$ leads to an electric current in the direction perpendicular to the field and the wave vector of the field-inducing perturbation. We dubbed this phenomenon the anomalous pseudo-Hall effect. The response of the chiral or valley current to the pseudoelectric field is also found to be nontrivial. Since the wave vector for $\mathbf{E}_5$ cannot be neglected, the frequency profile of the chiral conductivity is drastically different from its electric counterpart showing a step-like feature instead of a smooth Drude peak. The proposed effects can be investigated by driving sound waves in Weyl semimetals with broken time-reversal symmetry.
\end{abstract}

\maketitle

\section{Introduction}
\label{sec:Introduction}

Current response to various external perturbations, such as electromagnetic fields, is a time-honed probe of different material properties. In particular, the scaling of the optical conductivity with frequency and activation frequencies for the interband transitions provide an invaluable information about the band structure and the Fermi surface of materials.

At present, there is a surge of interest in novel topological phases of matter including quantum materials with a nontrivial relativistic-like energy spectrum known as Weyl and Dirac materials~\cite{Franz:book-2013,Wehling-Balatsky:rev-2014,Felser:rev-2017,Hasan-Huang:rev-2017,Burkov:rev-2018,Armitage:rev-2018,GMSS:book}. The band structure of these materials contains a number of band touching points known as Weyl nodes and Dirac points. Each of the Weyl nodes is characterized by a nontrivial topological charge whose sign is proportional to the chirality of the node. Unlike Dirac materials where a single Dirac node is allowed, Weyl nodes always appear in pairs of opposite chiralities~\cite{Nielsen:1981a,Nielsen:1981b,Nielsen:1981c}. The presence of several nodes or, equivalently, valleys allows not only for the electric but for chiral (valley) response connected with different charge densities or currents in nodes with different topological charges.

Optical conductivity in Weyl semimetals was theoretically studied in, e.g., Refs.~\cite{Ashby-Carbotte:2013,Ashby-Carbotte:2014-opt,Tabert-Carbotte:2016,Tabert-Carbotte:2016b,Carbotte:2016-tilt,Ahn-Min:2017,Roy-Juricic:2017,Burkov:2018-OptCond,Chen-Belyanin:2019,Sharma-Maslov:2021}. Among the characteristic features of the optical conductivity of Weyl materials, we mention the linear scaling with the frequency $\omega$ of the interband part of the conductivity $\sigma^{\text{inter}} \sim \omega$ if the frequency is large enough to overcome the Pauli blocking $\omega>2\mu$, the square-root spacing with a magnetic field of peaks corresponding to the Landau levels in the magneto-optical conductivity~\cite{Ashby-Carbotte:2014-opt}, the optical activity effects related to the anomalous Hall effect (AHE)~\cite{Yang-Lu:2011,Burkov-Balents:2011,Burkov-Hook:2011} such as the anomalous Kerr and Faraday effects~\cite{Kargarian-Trivedi:2015,Ahn-Min:2017,Chen-Belyanin:2019}, anomalous transmission and reflection of electromagnetic waves~\cite{Sukhachov-Glazman:2021-skin,Matus-Surowka:2021,Nandy-Pesin:2021}, etc.
Experimentally, the optical response and conductivity were investigated in Dirac~\cite{Chen-Wang-ZrTe5:2015,Jenkins-Drew:2016,Neubauer-Pronin:2016,Crassee-Akrap:2017-Cd3As2,Uykur-Kuntscher:2017,Cheng-Armitage:2019} and Weyl~\cite{Xu-Qiu:2016-TaAs-optics,Kimura-Felser:2017,Levy-Jenkins:2018,Cheng-Armitage:2019,Polatkan-Orlita:2019,Zu-Gopal:2020-TaAs-optics} semimetals; see also Refs.~\cite{Crassee-Orlita:2018,Pronin-Dressel:2020} for reviews.

One of the unique characteristics of Weyl and Dirac materials is that perturbations of their band structure, e.g., caused by mechanical deformations, could induce pseudo-electromagnetic (equivalently, axial) fields~\cite{Suzuura-Ando:2002,Vozmediano-Guinea:2010} (see Ref.~\cite{Ilan-Pikulin:rev-2019} for a review). Unlike conventional electromagnetic fields, the pseudo-electromagnetic ones act with the opposite sign on quasiparticles of opposite chiralities providing an efficient tool to induce the chiral response in Weyl materials. In the case of Dirac semimetals, there are proposals to generate the axial gauge fields by strains~\cite{Suzuura-Ando:2002,Vozmediano-Guinea:2010,Zhou-Shi:2013,Zubkov:2015,Cortijo-Vozmediano:2015,Cortijo-Vozmediano:2016,Pikulin-Franz:2016,Grushin:2016} as well as magnetization textures~\cite{Liu-Qiu:2012,Araki:rev-2020}. The first steps towards controlled realization of strains in Dirac semimetals were made in Ref.~\cite{Diaz-Moll:2022}. In graphene, pseudomagnetic fields were observed experimentally in Refs.~\cite{Levy-Crommie:2010,Liu-Loh:2018,Nigge-Damascelli:2019,Hsu-Yeh:2020}. Finally, in metamaterials whose energy spectrum resembles that of Weyl semimetals, pseudomagnetic fields were realized in Refs.~\cite{Roy-Grushin:2018,Peri-Huber:2019,Jia-Zhang:2019}.

Although pseudo-electromagnetic fields are distinct from the electromagnetic ones, one can still use a wide array of methods and tools developed for studying the response to electromagnetic waves to investigate the response of Weyl materials to pseudo-electromagnetic fields. Among the results derived in this approach, we could mention the manifestation of pseudomagnetic fields in collective modes~\cite{Gorbar:2016ygi,Gorbar:2016sey,Gorbar:2016vvg,Chernodub-Vozmediano:2019,Bugaiko-Sukhachov:2020}, strain-induced chiral magnetic effect and the negative pseudomagnetic resistivity~\cite{Cortijo-Vozmediano:2016,Grushin:2016,Pikulin-Franz:2016,Chernodub-Zubkov:2017,Huang:2017}, axial analogs of Hall responses~\cite{Huang:2017,Ghosh-Taraphder:2019}, the lensing of Weyl quasiparticles~\cite{Gorbar:2017dtp,Weststrom-Ojanen:2017,Soto-Garrido-Munoz:2018}, the acoustogalvanic effect~\cite{Sukhachov-Rostami:2019}, the axial magnetoelectric effect~\cite{Liang-Balatsky:2020}, etc.
In two-dimensional (2D) Dirac materials, dynamical strain-induced pseudo-electromagnetic fields might result in Hall currents~\cite{Oppen-Mariani:2009,Vaezi-Vozmediano:2013,Sela-Shalom:2019} and lead to an acoustoelectric effect~\cite{Hernandez-Minguez-Santos:2018,Bhalla-Rostami:2022}.

In this work, we study the response to external dynamical pseudo-electromagnetic fields in the Kubo linear response approach and determine the pseudoelectric conductivity in Weyl materials. This extends the previous studies by two of us in Refs.~\cite{Gorbar:2017wpi,Gorbar:2017yjn}. Unlike the case of optical response to ordinary electromagnetic fields, the wave vector of the pseudo-electromagnetic fields cannot be neglected. The reason is that the speed, for example, of sound waves, which can induce pseudo-electromagnetic fields, is much smaller than the speed of light. Among our main results, we mention a nontrivial Hall-like electric current response to a pseudoelectric field. Since this response requires a dynamical pseudoelectric field and has a form similar to the anomalous Hall effect, we dubbed it the anomalous pseudo-Hall effect (APHE). Furthermore, we derive the chiral conductivity tensor taking into account the wave vector of the external excitation $\mathbf{q}\neq\mathbf{0}$. The latter drastically affects the dependence of the chiral pseudoelectric conductivity on the frequency of the perturbations (e.g., sound) and makes the chiral response different from the conventional electric conductivity. In particular, the Drude peak has a finite width with a step-like feature even in the clean limit. Our results might be important for the development of the budding field of valley- and chiraltronics where the chirality of quasiparticles plays a prominent role.

The paper is organized as follows. In Sec.~\ref{sec:electric}, we introduce the model and study the electric current response to pseudo-electromagnetic fields. Section~\ref{sec:chiral} is devoted to the chiral response. Estimates of the proposed effects are provided in Sec.~\ref{sec:Estimates}. The results are discussed and summarized in Sec.~\ref{sec:Summary}. Technical details are given in a few appendixes at the end of the paper. Throughout this study we set $k_B=1$.

\section{Electric current response}
\label{sec:electric}

In this Section, we calculate the electric current response to the dynamical pseudoelectric field. The latter can be generated by dynamic strains such as sound or dynamical magnetization.

\subsection{Model and key notations}
\label{sec:electric-Model}

For simplicity, we employ a minimal linearized model of a Weyl material with broken time-reversal symmetry. The corresponding Hamiltonian is $\hat{H}=\text{diag}{\left( \hat{H}_{+}, \hat{H}_{-} \right)}$, where
\begin{equation}
\label{model-H-def}
\hat{H}_{\lambda}=\lambda \hbar v_{F} \bm{\sigma}\cdot \left(\mathbf{k}-\lambda\mathbf{b}\right).
\end{equation}
Here, $v_F$ is the Fermi velocity, $\bm{\sigma}$ is the vector of the Pauli matrices acting in the pseudospin space, $\mathbf{k}$ is the wave vector, $\lambda=\pm$ is the chirality or topological charge of the Weyl node, and $\mathbf{b}$ is the separation between the Weyl nodes in momentum space (the chiral shift).

We study the electric current response to a pseudoelectric field via the Kubo formalism (see Appendix~\ref{sec:app-1} for details). The latter can be used because the axial gauge field $\mathbf{A}_5$ interacts with Weyl quasiparticles in a similar way to the electromagnetic vector potential $\mathbf{A}$, i.e., instead of $\mathbf{k} \to \mathbf{k} + e \mathbf{A}/c$, we have $\mathbf{k} \to \mathbf{k} + \lambda e \mathbf{A}_5/c$. As one can see, the key difference is the dependence on chirality $\lambda$. Further, as the electromagnetic vector potential, the axial field can be time- and coordinate-dependent. The pseudoelectric field $\mathbf{E}_5$ is defined is the same way as the electric field $\mathbf{E}$, i.e., $\mathbf{E}_5(t,\mathbf{r}) = -\partial_t\mathbf{A}_5(t,\mathbf{r})/c$. Therefore,  $\mathbf{E}_5(\Omega,\mathbf{q}) =i\Omega \mathbf{A}_5(\Omega,\mathbf{q})/(\hbar c)$ in the frequency-momentum space. Notice that similarly to the Weyl node separation vector $\mathbf{b}$, the axial gauge field breaks the time-reversal symmetry but preserves the inversion one. Therefore, the pseudoelectric field preserves both the time-reversal and inversion symmetries.

In order to avoid confusion, we use the superscript $\rm (p)$ for the electric current response the superscript $\rm (5,p)$ for the chiral current response to the pseudoelectric field. The real part of the electric conductivity tensor reads
\begin{equation}
\label{model-sigma-ij-def}
\text{Re}\,\sigma_{ij}^{\rm (p)}(\Omega,\mathbf{q}) = \hbar \frac{ \mbox{Im}\, \Pi_{ij}^{\rm (p)}(\Omega+i0,\mathbf{q})}{\Omega},
\end{equation}
where the correlation function of the electric and chiral currents is
\begin{eqnarray}
\label{model-Pi-ij-def}
\Pi_{ij}^{\rm (p)}(\Omega+i0,\mathbf{q})\! &=&\! e^2 \int_{-\infty}^{+\infty}\!d\omega \!\int_{-\infty}^{+\infty}\!d\omega^{\prime} \frac{f(\omega)-f(\omega^{\prime})} {\omega^{\prime} -\omega-\Omega-i0}\nonumber \\
&\times& \!\! \int\! \frac {d^3 k}{(2 \pi)^3} \text{Tr}\!\left[\hat{v}_i \hat{A}(\mathbf{k},\omega) \gamma^5 \hat{v}_j \hat{A}(\mathbf{k}+\mathbf{q},\omega^{\prime})\right]. \nonumber\\
\end{eqnarray}
Here, $f(\omega)=1/\left[e^{\left(\omega - \mu\right)/T}+1\right]$ is the Fermi-Dirac distribution function, $\mu$ is the chemical potential, $T$ is temperature, $\hat{v}_i = \partial_{k_i} \hat{H}/\hbar$ is the velocity operator, $\gamma^5=\text{diag}{\left(\hat{I}, -\hat{I}\right)}$ is the chirality matrix, and $\hat{A}(\omega;\mathbf{k})$ is the spectral function; see Appendix~\ref{sec:app-1} for its definition. In what follows, we mostly focus on the case of small temperature compared to the Fermi energy, i.e., set $T\to0$. This allows us to obtain simpler analytical expressions without neglecting any important features.
Finally, we notice that, unlike the conventional electric conductivity, we do not consider the uniform limit $q\to0$ in the response to the pseudoelectric field.

For Hamiltonian (\ref{model-H-def}), the spectral function $\hat{A}(\omega;\mathbf{k})=\text{diag}{\left( \hat{A}_{+}(\omega;\mathbf{k}), \hat{A}_{-}(\omega;\mathbf{k}) \right)}$ is a diagonal matrix where $\hat{A}_{\lambda}$ with $\lambda=\pm$ reads
\begin{eqnarray}
\label{model-spectral-function-def}
\hat{A}_{\lambda}(\omega;\mathbf{k}) &=& \frac{1}{2}\left[\delta(\omega - \varepsilon_{\mathbf{k},\lambda}) +\delta(\omega + \varepsilon_{\mathbf{k},\lambda})\right] \hat{I} \nonumber\\
&+& \frac{\hat{H}_{\lambda}}{2\varepsilon_{\mathbf{k},\lambda}} \left[\delta(\omega - \varepsilon_{\mathbf{k},\lambda}) -\delta(\omega + \varepsilon_{\mathbf{k},\lambda})\right].
\end{eqnarray}
Here, $\delta(x)$ is the $\delta$-function and $\varepsilon_{\mathbf{k},\lambda}=\hbar v_F |\mathbf{k}- \lambda \mathbf{b}|$ is the dispersion relation of quasiparticles with chirality $\lambda$.

\subsection{Conductivity tensor}
\label{sec:electric-sigma}

Let us calculate the pseudoelectric conductivity tensor. By using the Sokhotski-Plemelj theorem, we rewrite Eq.~(\ref{model-sigma-ij-def}) as
\begin{widetext}
\begin{eqnarray}
\label{el-sigma-ij-def}
\text{Re}\,\sigma_{ij}^{\rm (p)}(\Omega,\mathbf{q})&=&\frac {\hbar e^2 }{\Omega } \text{v.p.} \int_{-\infty}^{+\infty} d \omega^{\prime} \int_{-\infty}^{+\infty}d\omega \frac{f(\omega)-f(\omega^{\prime})}{\omega^{\prime}-\omega-\Omega} \int \frac{d^3 k}{(2\pi)^3} \mbox{Im} \left\{\text{Tr}\left[\hat{v}_i \hat{A}(\mathbf{k},\omega)\gamma^5 \hat{v}_j \hat{A}(\mathbf{k}+\mathbf{q},\omega^{\prime})\right] \right\} \nonumber\\
&+& \frac{\pi \hbar e^2}{\Omega} \int_{-\infty}^{+\infty}d\omega \left[f(\omega)-f(\omega+\Omega)\right] \int \frac{d^3 k}{(2\pi)^3}
\mbox{Re} \left\{\text{Tr}\left[\hat{v}_i \hat{A}(\mathbf{k},\omega) \gamma^5 \hat{v}_j  \hat{A}(\mathbf{k}+\mathbf{q},\omega+\Omega)\right] \right\},
\end{eqnarray}
\end{widetext}
where $\text{v.p.}$ denotes the principal value and we integrated over $\omega^{\prime}$ in the second term.

After straightforward but lengthy calculations, it can be shown that the only nonzero components of the conductivity tensor (\ref{el-sigma-ij-def}) are $\text{Re}\,\sigma_{ij}^{\rm (p)}(\Omega,\mathbf{q})=-\text{Re}\,\sigma_{ji}^{\rm (p)}(\Omega,\mathbf{q})$ with $i\neq j$. Other components vanish after the summation over Weyl nodes. As we show in Appendix~\ref{sec:app-3}, the conventional expressions for the electric conductivity tensor are reproduced if one omits $\gamma^{5}$ in Eq.~(\ref{el-sigma-ij-def}). The off-diagonal components of the conductivity tensor are determined only by the first term in Eq.~(\ref{el-sigma-ij-def}); see Appendix~\ref{sec:app-2} for the details of the derivation. The result reads as
\begin{widetext}
\begin{eqnarray}
\label{el-sigma-xy-mu-neq-0-Main}
\text{Re}\, \sigma_{ij}^{\rm (p)}(\Omega,\mathbf{q}) &=& \varepsilon_{ijl} \frac{e^2 q_l}{4 \pi^2 \hbar} \Bigg \{1-\frac{\Omega \left(\tilde{q}^2-\Omega^2\right)}{8  \tilde{q} ^3} \Bigg[ 4 \mu \Omega  \ln \left|\frac{\tilde{q}^2}{(\tilde{q}-\Omega)^2} \frac{ (\tilde{q} - \Omega +2  \mu ) (\tilde{q} - \Omega -2  \mu )}{ (\tilde{q} + \Omega -2  \mu) ( \tilde{q} + \Omega + 2  \mu)}\right| -8 \mu \tilde{q}
\nonumber\\
&+& \left(\tilde{q}^2-\Omega^2 -4 \mu^2 \right) \ln \left| \frac{(\tilde{q} + \Omega +2 \mu)(\tilde{q} -\Omega +2 \mu)}{(\tilde{q} + \Omega-2 \mu)(\tilde{q} - \Omega -2 \mu)}\right| \Bigg] \Bigg\} =\varepsilon_{ijl} \frac{e^2 q_l}{4 \pi^2 \hbar} \left[1+ \frac{\mu}{\Omega}\left(1+\frac{8}{3} \frac{\mu^2}{\Omega^2} \right)\frac{v_s^2}{v_F^2} +o \left(\frac{v_s^2}{v_F^2}\right)  \right], \nonumber\\
\end{eqnarray}
\end{widetext}
where $\tilde{q}=\hbar v_F q$. In the last expression, we used the sound dispersion relation $\Omega=\hbar v_s q$, expanded in small $v_s/v_F$, and assumed that $\mu/\Omega \lesssim v_F/(2v_s)$.

The off-diagonal conductivity $\text{Re}\,\sigma_{ij}^{(p)}$ is not zero even at $\mu=0$ and acquires a simple form
\begin{equation}
\label{el-sigma-xy-mu=0-fin}
\text{Re}\, \sigma_{ij}^{\rm (p)}(\Omega,\mathbf{q}) =\varepsilon_{ijl}\frac{e^2 q_l}{4 \pi^2 \hbar}.
\end{equation}
Notice that there is no dependence on the chiral shift $\mathbf{b}$ and, at $\mu=0$, on $\Omega$.

Let us discuss the physical meaning of the results given in Eqs.~(\ref{el-sigma-xy-mu-neq-0-Main}) and (\ref{el-sigma-xy-mu=0-fin}). In essence, the obtained conductivity tensor means that there is a Hall-like electric current induced by the pseudoelectric field in the direction perpendicular to the field, i.e., $J_i(\Omega,\mathbf{q})=\sigma_{ij}^{\rm (p)}(\Omega,\mathbf{q}) E_{5,j}(\Omega,\mathbf{q})$. Here, the pseudoelectric field is $\mathbf{E}_5(\Omega,\mathbf{q}) = i\Omega \mathbf{A}_5(\Omega,\mathbf{q})/(\hbar c)$. The obtained response resembles the AHE with $\text{Re}\, \sigma_{ij} = - \varepsilon_{ijl} e^2b_l/(2\pi^2 \hbar)$ but it is driven by the dynamical pseudoelectric rather than electric field. Since the pseudoelectric field originates from a dynamical chiral shift, see also the discussion in Sec.~\ref{sec:Estimates}, we dub the proposed effect the anomalous pseudo-Hall effect.

Heuristically, one can indeed expect the APHE conductivity $\sigma_{\rm APHE}^{(p)} \propto q$, where $\text{Re}\, \sigma_{ij}^{(p)} =\epsilon_{ijl} q_l\sigma_{\text{APHE}}^{(p)}/q$, in the response to the pseudoelectric field on similar grounds as one expects the appearance of the AHE conductivity $\sigma_{\rm AHE} \propto b$ in the response to an electric field. Indeed, momentum and chiral shift enter Hamiltonian (\ref{model-H-def}) and spectral function (\ref{model-spectral-function-def}) with different chirality prefactor $\lambda$. After summing over all Weyl nodes, the contributions proportional to $\mathbf{q}$ in the AHE cancel. On the other hand, the pseudoelectric field $\mathbf{E}_5$ in the APHE provides the additional factor $\lambda$ that allows for a nontrivial response proportional to the momentum $\mathbf{q}$.
We notice that, unlike the Hall response in pseudolectric and pseudomagnetic fields discussed in Ref.~\cite{Sela-Shalom:2019}, no external pseudomagnetic field is needed for the APHE.

\begin{figure}[t]
\centering
\includegraphics[width=0.45\textwidth]{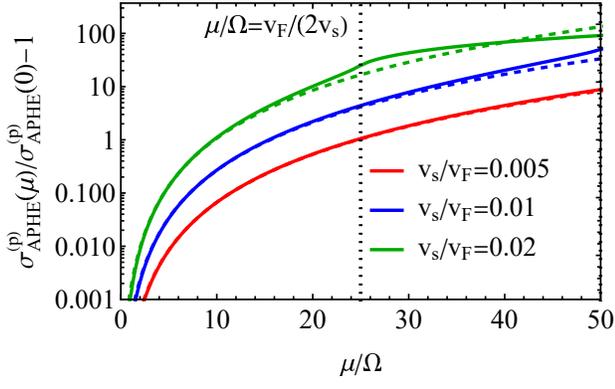}
\vspace{-0.05cm}
\caption{The dependence of the relative APHE conductivity $\sigma_{\text{APHE}}^{\rm (p)}(\mu)/\sigma_{\text{APHE}}^{\rm (p)}(0)-1$ on $\mu/\Omega$ for a few values of $v_s/v_F$. Solid and dashed lines correspond to the exact and approximate results in Eq.~(\ref{el-sigma-xy-mu-neq-0-Main}). The vertical dotted line denotes the limit of applicability of the approximate expression given in the last line in (\ref{el-sigma-xy-mu-neq-0-Main}), i.e., $\mu/\Omega \lesssim v_F/(2v_s)$. We used the sound dispersion relation $\Omega =\hbar v_s q$ and defined $\text{Re}\, \sigma_{ij}^{(p)} =\epsilon_{ijl} q_l\sigma_{\text{APHE}}^{(p)}/q$.
}
\label{fig:Conductivity_xy_mu}
\end{figure}

We present the dependence of the relative APHE conductivity $\sigma_{\text{APHE}}^{\rm (p)}(\mu)/\sigma_{\text{APHE}}^{\rm (p)}(0)-1$ on $\mu$ in Fig.~\ref{fig:Conductivity_xy_mu}. As one can see, deviations from the result in Eq.~(\ref{el-sigma-xy-mu=0-fin}) appear only for large Fermi energies $\mu \gtrsim\Omega$; see solid and dashed lines in Fig.~\ref{fig:Conductivity_xy_mu}.
Since the Fermi energy in typical Weyl semimetals is usually larger than the frequency $\Omega$, we expect the corresponding matter contributions~\footnote{The matter contribution is defined as the part of the conductivity connected with the Fermi energy without the topological part, i.e., $\sigma_{\text{APHE}}^{\rm (p)}(\mu)-\sigma_{\text{APHE}}^{\rm (p)}(0)$.} to play an important role in the APHE. Indeed, as one can see from Fig.~\ref{fig:Conductivity_xy_mu}, the effect of the Fermi energy is negligible only for small sound velocities $v_s\ll v_F$ or small Fermi energies $\mu \ll \Omega$. Therefore, in order to observe the topological contribution given in Eq.~(\ref{el-sigma-xy-mu=0-fin}), semimetals with small Fermi energy should be used.

\section{Chiral current response}
\label{sec:chiral}

In this Section, we study the chiral current response to the dynamical pseudoelectric field. The corresponding conductivity tensor has the form~(\ref{model-sigma-ij-def}) with the replacement $\Pi_{ij}^{\rm (p)}(\Omega+i0,\mathbf{q}) \to \Pi_{ij}^{\rm (5,p)}(\Omega+i0,\mathbf{q})$, where $\Pi_{ij}^{\rm (5,p)}(\Omega+i0,\mathbf{q})$ is obtained by introducing $\gamma^5 \hat{v}_i$ instead of $\hat{v}_i$ in Eq.~(\ref{model-Pi-ij-def}).

On the symmetry grounds, one could expect that the chiral current response to the pseudoelectric field is similar to the electric current response to the electric field. This is indeed the case if there are no terms in the Hamiltonian that intermix chiralities. Then, the chiral conductivity tensor $\sigma_{ij}^{\rm (5,p)}(\Omega,\mathbf{q})$ has the same form as the conventional electric conductivity tensor $\sigma_{ij}(\Omega,\mathbf{q})$. However, since the speed of dynamical perturbations such as sound or magnetization waves is much smaller than the speed of light, one can no longer neglect the wave vector $\mathbf{q}$.

The details of the calculations and intermediate expressions are given in Appendix~\ref{sec:app-3}. In what follows, we present the key results assuming  vanishing temperature. As we will see below, the diagonal (dissipative) components of the chiral conductivity tensor are nonvanishing. Therefore, it is important to include the effects of disorder. However, to emphasize the role of the wave vector, we first consider the clean limit where the analytical results for the chiral conductivity tensor have a simple form; see Sec.~\ref{sec:chiral-clean}. Then, we present the numerical results for two models of disorder in Sec.~\ref{sec:chiral-Gamma}.

\subsection{Clean limit}
\label{sec:chiral-clean}

Let us start with the clean limit. We find it convenient to separate the interband and intraband parts of the diagonal conductivity tensor in the directions perpendicular $\sigma_{\perp}^{\rm (5,p)}(\Omega,\mathbf{q})$ and parallel $\sigma_{\parallel}^{\rm (5,p)}(\Omega,\mathbf{q})$ to the wave vector $\mathbf{q}$~\footnote{For example, for the wave vector along the $z$-axis$, \mathbf{q}\parallel \hat{\mathbf{z}}$, we have $\sigma_{\perp}^{\rm (5,p)}(\Omega,\mathbf{q}) = \sigma_{xx}^{\rm (5,p)}(\Omega,\mathbf{q}) =\sigma_{yy}^{\rm (5,p)}(\Omega,\mathbf{q})$ and $\sigma_{\parallel}^{\rm (5,p)}(\Omega,\mathbf{q})= \sigma_{zz}^{\rm (5,p)}(\Omega,\mathbf{q})$}.
For the interband part of the conductivity tensor, we obtain
\begin{widetext}
\begin{eqnarray}
\label{chiral-sigma-xx-IB}
\text{Re}\,\sigma_{\perp}^{\rm (5,p), \text{inter}}(\Omega,\mathbf{q}) &=& \frac{e^2}{12 \pi \hbar^2 v_F} \frac{\Omega^2-\tilde{q}^2}{\Omega}\Theta\left(\Omega -\tilde{q}\right)
\Bigg\{
\left[1- \Theta\left(\mu -\tilde{q}\right) \Theta\left(2\mu -\tilde{q} -\Omega\right)\right] \Theta\left(2\mu + \tilde{q} -\Omega\right) \frac{\tilde{q}+\Omega-2\mu}{8 \tilde{q}^3} \nonumber\\
&\times& \left[(\Omega-2 \mu)^2-\tilde{q}(\Omega-2\mu)+4\tilde{q}^2\right] +\Theta\left(\Omega - \tilde{q} -2\mu\right)
\Bigg\}
\end{eqnarray}
\end{widetext}
and
\begin{widetext}
\begin{eqnarray}
\label{chiral-sigma-zz-IB}
\text{Re}\,\sigma_{\parallel}^{(5,p), \text{inter}}(\Omega,\mathbf{q}) &=&
\frac{e^2}{12\pi \hbar^2 v_F} \Omega \, \Theta\left(\Omega -\tilde{q}\right) \Bigg\{ \left[1- \Theta\left(\mu -\tilde{q}\right) \Theta\left(2\mu - \tilde{q} -\Omega \right)\right] \Theta\left(2\mu + \tilde{q} -\Omega \right) \frac{\left(2\mu+2 \tilde{q}-\Omega\right) \left(2\mu -\Omega-\tilde{q}\right)^2}{4 \tilde{q}^3} \nonumber\\
&+&\Theta\left(\Omega - \tilde{q} -2\mu\right)\Bigg\},
\end{eqnarray}
\end{widetext}
where $\Theta(x)$ is a step function and we assumed that $\mu \geq 0$. In the case of sound waves, we use the dispersion relation $q=\Omega/\left(\hbar v_s\right)$ with $v_s$ being the sound velocity. In this case, however, the interband contribution to the pseudoelectric conductivity is suppressed. Indeed, since $v_F\gg v_s$ in (semi)metals, $\Theta\left(\Omega -\tilde{q}\right) =0$. On the other hand, the interband parts of the pseudoelectric conductivity might be nonzero in metamaterials where the effective Fermi velocity is much smaller~\cite{Peri-Huber:2019}; see also Appendix~\ref{sec:app-3-Re} for results at $v_s>v_F$.

Next, let us discuss the intraband (or Drude) part of the chiral pseudoelectric conductivity tensor. We derive
\begin{widetext}
\begin{eqnarray}
\label{chiral-sigma-xx-D}
\text{Re}\,\sigma_{\perp}^{(5,p), \text{intra}}(\Omega,\mathbf{q}) &=& \frac{e^2 \left(\tilde{q}^2-\Omega^2 \right)}{96 \pi \hbar^2 v_F \Omega \tilde{q}^3}
\Bigg\{\Theta(\tilde{q} -\Omega)\Theta(2\mu -\tilde{q} +\Omega) \left(2\mu+\Omega- \tilde{q}\right) \left[\left(2\mu +\Omega\right) \left(2\mu +\Omega +\tilde{q}\right) +4\tilde{q}^2\right] \nonumber\\
&-& \left[\Theta(\tilde{q}-\mu)\Theta (2\mu-\tilde{q})\Theta(2\mu-\tilde{q}-\Omega) + \Theta(\tilde{q}-\Omega) \Theta(\mu-\tilde{q}) \right] \left(2\mu -\Omega -\tilde{q}\right) \nonumber\\
&\times& \left[\left(2\mu -\Omega\right) \left(2\mu -\Omega +\tilde{q}\right) +4\tilde{q}^2\right]
\Bigg\}
\end{eqnarray}
\end{widetext}
and
\begin{widetext}
\begin{eqnarray}
\label{chiral-sigma-zz-D}
\text{Re}\,\sigma_{\parallel}^{(5,p), \text{intra}}(\Omega,\mathbf{q}) &=&
\frac{e^2\Omega}{48 \pi \hbar^2 v_F \tilde{q}^3}
\Bigg\{ \Theta\left(\tilde{q} -\Omega\right)\Theta\left(2\mu-\tilde{q} +\Omega\right) \left(2\mu +\Omega -\tilde{q}\right)^2 \left(2\mu+\Omega +2\tilde{q}\right) \nonumber\\
&-& \left[\Theta(\tilde{q}-\mu)\Theta(2\mu-\tilde{q})\Theta(2\mu-\tilde{q}-\Omega) +\Theta(\tilde{q}-\Omega) \Theta(\mu-\tilde{q})\right] \left(2\mu -\Omega +2\tilde{q}\right) \left(2\mu -\Omega -\tilde{q}\right)^2
\Bigg\}.
\end{eqnarray}
\end{widetext}
It is interesting to notice that the corresponding intraband parts of the conductivity tensor no longer contain $\delta(\Omega)$ (i.e., the Drude peak) for $q\neq0$. We checked that the conventional expression with $\text{Re}\,\sigma_{\parallel}^{(5,p), \text{intra}}(\Omega,\mathbf{0})\propto \delta(\Omega)$ is restored in the limit $q\to0$; see Appendix~\ref{sec:app-3} for details.

\begin{figure*}[t]
\centering
\subfigure[]{\includegraphics[width=0.45\textwidth]{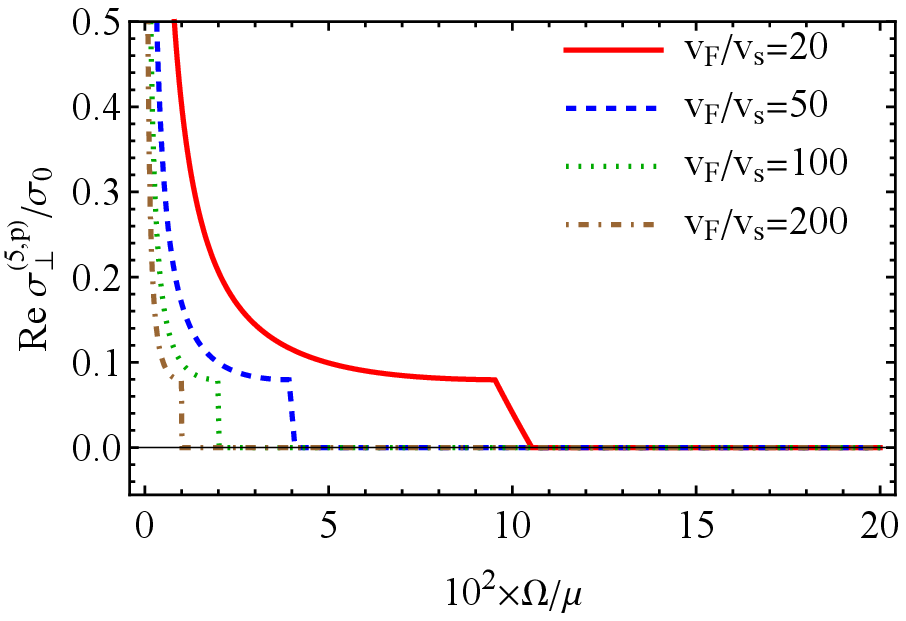}}
\hspace{0.01\textwidth}
\subfigure[]{\includegraphics[width=0.45\textwidth]{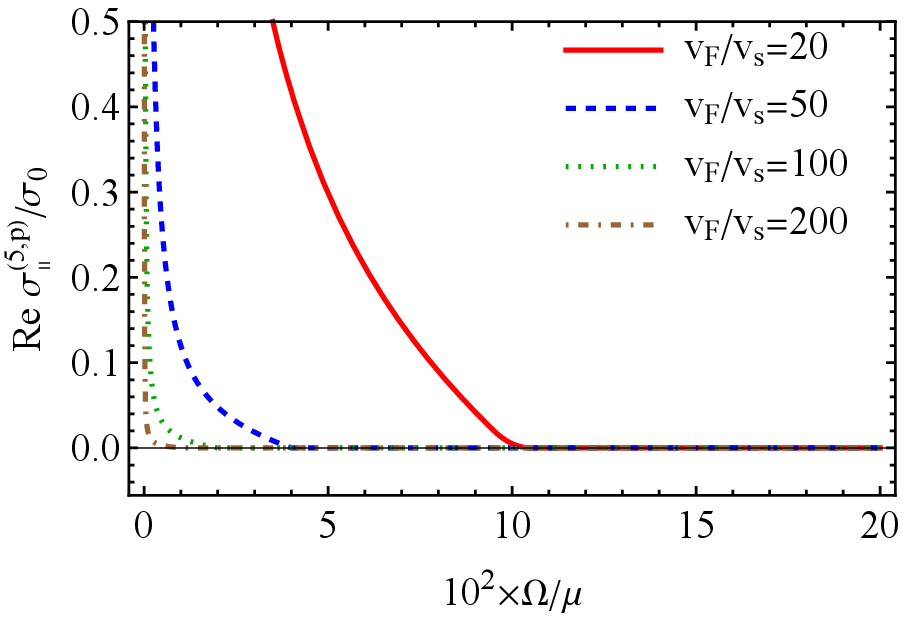}}
\vspace{-0.05cm}
\caption{The dependence of the diagonal components of the chiral pseudoelectric conductivity on $\Omega/\mu$ for a few values of $v_F/v_s$. Panels (a) and (b) show $\text{Re}\,\sigma_{\perp}^{\rm (5,p)}(\Omega, \mathbf{q})$ and $\text{Re}\,\sigma_{\parallel}^{\rm (5,p)}(\Omega, \mathbf{q})$, respectively. We used $\sigma_0=e^2\mu/(\hbar^2 v_F)$. Notice that the intraband conductivity tensor components have a conventional Drude peak at $\Omega\to0$ for $v_F<v_s$. The interband components vanish for $v_F>v_s$.
}
\label{fig:Conductivity_intra_nonzero_q_T=0}
\end{figure*}

The dependence of $\sigma_{\perp}^{(5,p), \text{intra}}(\Omega,\mathbf{q})$ and $\sigma_{\parallel}^{(5,p), \text{intra}}(\Omega,\mathbf{q})$ on $\Omega$ is shown in Fig.~\ref{fig:Conductivity_intra_nonzero_q_T=0} for a few values of $v_F/v_s$. Notice also that unlike the interband part of the conductivity, the nontrivial intraband one requires $v_F\geq v_s$. The Drude-peak behavior is gradually restored as $v_F/v_s$ increases. As one can see from Fig.~\ref{fig:Conductivity_intra_nonzero_q_T=0}, there are two regions with different dependence on $\Omega$. Indeed, the first term in the curly brackets in Eqs.~(\ref{chiral-sigma-xx-D}) and (\ref{chiral-sigma-zz-D}) is nonzero for
\begin{equation}
\label{chiral-sigma-D-first}
\frac{\Omega}{\mu} \leq \frac{2v_s}{v_F-v_s}.
\end{equation}
The second term contributes for
\begin{equation}
\label{chiral-sigma-D-second}
\frac{\Omega}{\mu} \leq \frac{2v_s}{v_F+v_s}.
\end{equation}
Its contribution is manifested in a kink-like feature seen in Fig.~\ref{fig:Conductivity_intra_nonzero_q_T=0}(a). The change of the slope is also present in Fig.~\ref{fig:Conductivity_intra_nonzero_q_T=0}(b) but it is less evident.

It is important to emphasize also that the characteristic frequency scale at which the nontrival features of the chiral pseudoelectric conductivity are observed is determined by $v_s\Omega/(\mu v_F)$ rather than $\Omega/\mu$ as in the case of conventional optical conductivity tensor. This difference originates from the wave vector dependence of $\sigma_{ij}^{\rm (5,p)}(\Omega, \mathbf{q})$.

As to $\text{Re}\,\sigma_{\perp}^{\rm (5,p)}(\Omega, \mathbf{q})$ and $\text{Re}\,\sigma_{\parallel}^{\rm (5,p)}(\Omega, \mathbf{q})$, we notice that these components of the conductivity tensor are responsible for the dissipation of the pseudoelectric field similar for the Joule heating for an electric field. In the case of the sound-induced $\mathbf{E}_5$, the role of the vector potential in the sound attenuation was discussed in Ref.~\cite{Pesin-Ilan:2020} albeit in the semiclassical approach and with more emphasis on the role of the chiral anomaly.

Finally, let us discuss the off-diagonal components of the conductivity tensor. We obtain the expression that resembles the standard pseudo-AHE conductivity (see Appendix~\ref{sec:app-3} for details)
\begin{equation}
\label{chiral-sigma-xy}
\text{Re}\,\sigma_{ij}^{\rm (5,p)}(\Omega,\mathbf{q}) = -\text{Re}\,\sigma_{ji}^{\rm (5,p)}(\Omega,\mathbf{q}) = - \varepsilon_{ijl} \frac{e^2 b_l}{2 \pi^2 \hbar}.
\end{equation}
The chiral AHE conductivity does not depend on frequency $\Omega$ and wave vector $\mathbf{q}$ in the linearized model.

Until now, we focused on the real part of the chiral conductivity tensor. For the sake of completeness, let us also present the imaginary part. The full conductivity tensor accounting for both $\mu$ and $\mathbf{q}$ is cumbersome. We present its diagonal components in Appendix~\ref{sec:app-3}. For a simplified case $\mu=0$, the diagonal components of the chiral pseudoelectric conductivity tensor read:
\begin{eqnarray}
\label{chiral-sigma-Im-inter-mu=0}
\text{Im}\, \sigma_{\perp}^{(5,p), \text{inter}}(\Omega,\mathbf{q}) &=& -\frac{e^2}{12 \pi^2 \hbar^2 v_F \Omega} \bigg[2 \tilde{q}^2 \ln\left|\frac{\Lambda }{\tilde{q}}\right|\nonumber\\
&+& \left(\Omega ^2-\tilde{q}^2\right) \ln\left|\frac{\Lambda^2}{\Omega ^2-\tilde{q}^2}\right| \bigg], \\
\label{chiral-sigma-Im-intra-mu=0}
\text{Im}\, \sigma_{\parallel}^{(5,p), \text{inter}}(\Omega,\mathbf{q}) &=&- \frac{e^2 \Omega}{12 \pi^2 \hbar^2 v_F} \ln\left|\frac{\Lambda^2}{\Omega ^2-\tilde{q}^2}\right|,\\
\text{Im}\, \sigma_{\perp}^{(5,p), \text{intra}}(\Omega,\mathbf{q}) &=& \text{Im}\, \sigma_{\parallel}^{(5,p), \text{intra}}(\Omega,\mathbf{q}) =0.
\end{eqnarray}
Here, $\Lambda$ is the energy cutoff, whose physical meaning could be related to the applicability of the linear dispersion relation. The obtained imaginary part of the chiral pseudoelectric conductivity is the same as that of the optical conductivity in Weyl semimetals~\cite{Rosenstein:2013,Roy-DasSarma:2016}.

We plot the analytical results given at the end of Appendix~\ref{sec:app-3-Im} for $\mu\neq0$ in Fig.~\ref{fig:Conductivity_Im}. Notice that unlike the transverse component of the conductivity tensor $\text{Im}\, \sigma_{\perp}^{\rm (5,p)}(\Omega, \mathbf{q})$, the longitudinal component $\text{Im}\, \sigma_{\parallel}^{\rm (5,p)}(\Omega, \mathbf{q})$ has a pole at $\Omega=2\tilde{q}$. The latter corresponds to the step-like feature in Fig.~\ref{fig:Conductivity_intra_nonzero_q_T=0}(a). While the interband part of the conductivity diminishes with $\Omega$, the intraband one rises with frequency leading to a nonmonotonic behavior.

\begin{figure*}[t]
\centering
\subfigure[]{\includegraphics[width=0.45\textwidth]{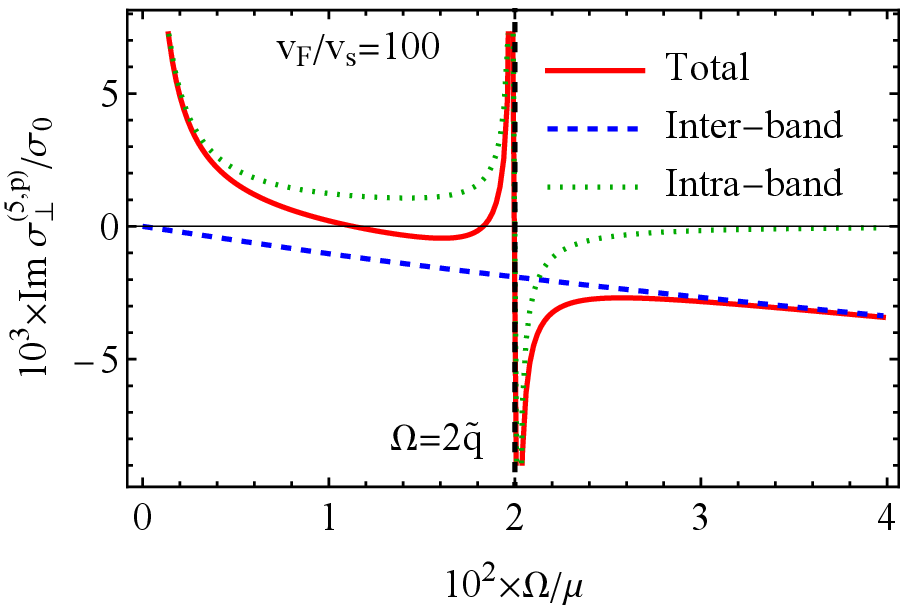}}
\hspace{0.01\textwidth}
\subfigure[]{\includegraphics[width=0.45\textwidth]{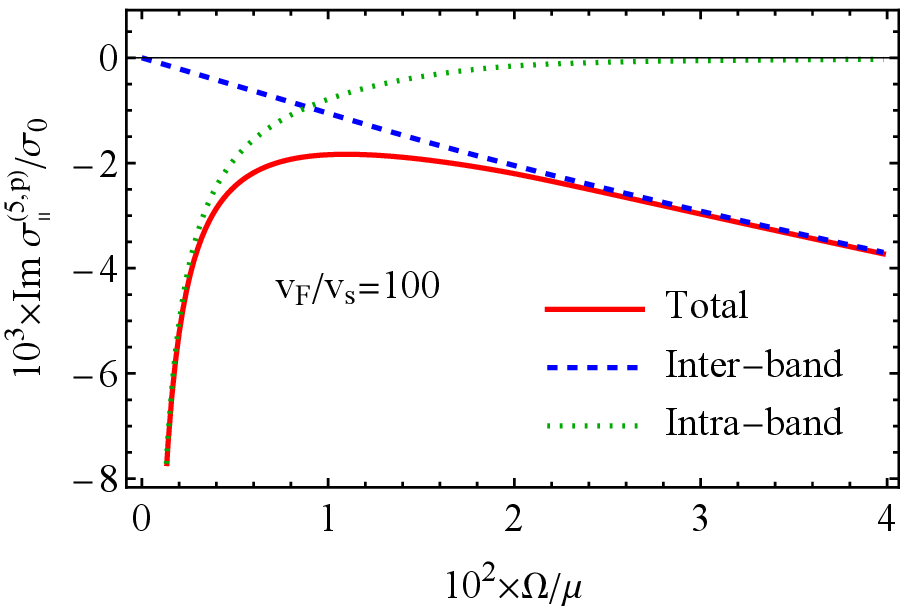}}
\vspace{-0.05cm}
\caption{The dependence of the diagonal components of the imaginary part of the chiral pseudoelectric conductivity on $\Omega/\mu$. Panels (a) and (b) show $\text{Im}\, \sigma_{\perp}^{\rm (5,p)}(\Omega, \mathbf{q})$ and $\text{Im}\, \sigma_{\parallel}^{\rm (5,p)}(\Omega, \mathbf{q})$, respectively. We used $\Omega=\hbar v_s q$, $\tilde{q}=\hbar v_F q$, and $\sigma_0=e^2\mu/(\hbar^2 v_F)$, as well as fixed $v_F/v_s=100$ and \revision{$\Lambda/\mu=10^3$} in both panels.
}
\label{fig:Conductivity_Im}
\end{figure*}

\subsection{Effects of disorder}
\label{sec:chiral-Gamma}

In this Section, we demonstrate that the key features of the chiral conductivity tensor survive in the presence of disorder. The latter is taken into account via the self-energy contributions that lead to the broadening of the $\delta$-functions in the spectral function given in Eq.~(\ref{model-spectral-function-def}),
\begin{equation}
\label{chiral-G-Lorentzian}
\delta\left(\omega -\varepsilon_{\mathbf{k},\lambda}\right) \to \delta_{\Gamma}\left(\omega -\varepsilon_{\mathbf{k},\lambda}\right)= \frac{1}{\pi}\frac{\Gamma(\omega)}{\left(\omega -\varepsilon_{\mathbf{k},\lambda}\right)^2+\Gamma^2(\omega)},
\end{equation}
see, also, Appendix~\ref{sec:app-1}. We consider two models of disorder. In the first model, we assume a constant broadening $\Gamma=\Gamma_0$. The second model has a frequency-dependent broadening $\Gamma(\omega) = \Gamma_1 \omega^2$~\cite{Burkov-Hook:2011}, which corresponds to local scatterers. Therefore, the total effective broadening can be defined as
\begin{equation}
\label{chiral-G-Gamma-tot}
\Gamma(\omega) = \Gamma_0 + \Gamma_1 \omega^2.
\end{equation}

\begin{figure*}[t]
\centering
\subfigure[]{\includegraphics[width=0.45\textwidth]{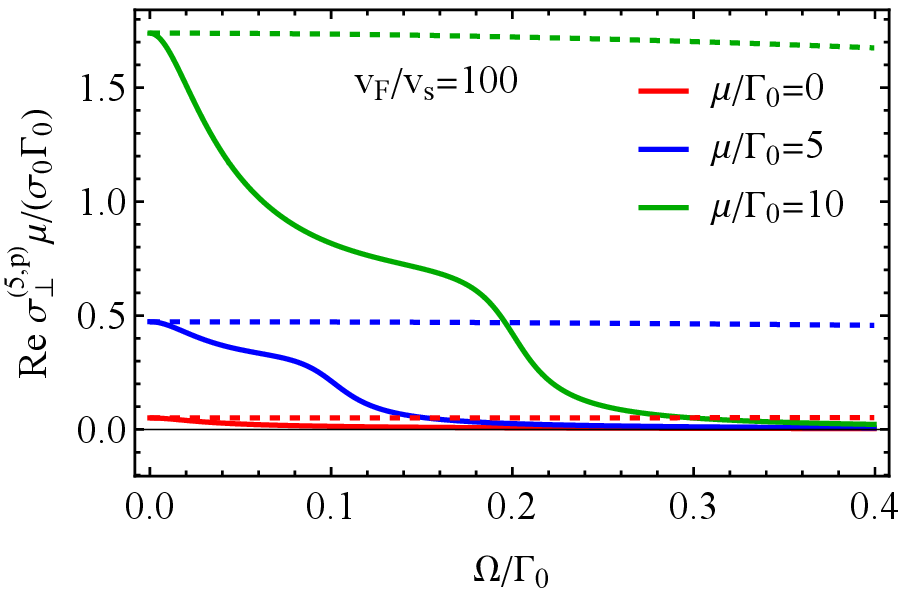}}
\hspace{0.01\textwidth}
\subfigure[]{\includegraphics[width=0.45\textwidth]{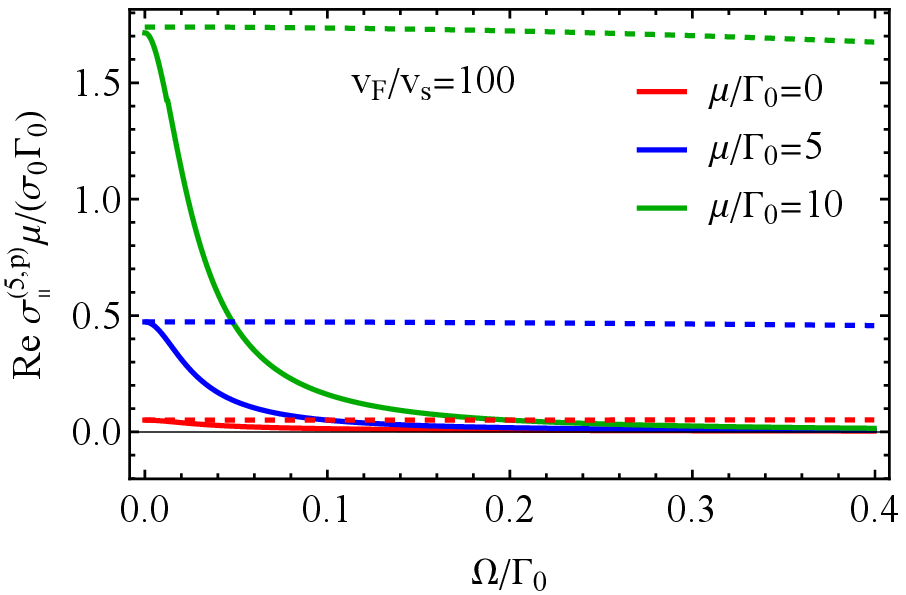}}
\vspace{-0.05cm}
\caption{The dependence of the diagonal components of the chiral pseudoelectric conductivity on $\Omega/\Gamma_0$ for a few values of $\mu/\Gamma_0$. Panels (a) and (b) show $\text{Re}\, \sigma_{\perp}^{\rm (5,p)}(\Omega, \mathbf{q})$ and $\text{Re}\,\sigma_{\parallel}^{\rm (5,p)}(\Omega, \mathbf{q})$, respectively. Solid lines correspond to the case $q=\Omega/(\hbar v_s)$ and dashed lines show the results at $q=0$. We used $\sigma_0=e^2\mu/(\hbar^2 v_F)$, $v_F/v_s=100$, and the broadening function (\ref{chiral-G-Gamma-tot}) with $\Gamma_1=0$.
}
\label{fig:Conductivity_Gamma_nonzero_q_T=0_Gamma0}
\end{figure*}

We show the dependence of the diagonal components of the chiral pseudoelectric conductivity on $\Omega/\Gamma_0$ for a simple model of disorder $\Gamma(\omega) = \Gamma_0$ in Fig.~\ref{fig:Conductivity_Gamma_nonzero_q_T=0_Gamma0}. As one can see from Fig.~\ref{fig:Conductivity_Gamma_nonzero_q_T=0_Gamma0}(a), the step-like feature determined by Eq.~(\ref{chiral-sigma-D-first}) remains clearly visible on top of the broadened profile. It might be difficult to distinguish the change of the frequency profile of the conductance for $v_F\gg v_s$, however. Nevertheless, the conductivity is qualitatively different at $q\neq 0$ and $q=0$, cf. Figs.~\ref{fig:Conductivity_Gamma_nonzero_q_T=0_Gamma0}(a) and \ref{fig:Conductivity_Gamma_nonzero_q_T=0_Gamma0}(b).

\begin{figure*}[t]
\centering
\subfigure[]{\includegraphics[width=0.45\textwidth]{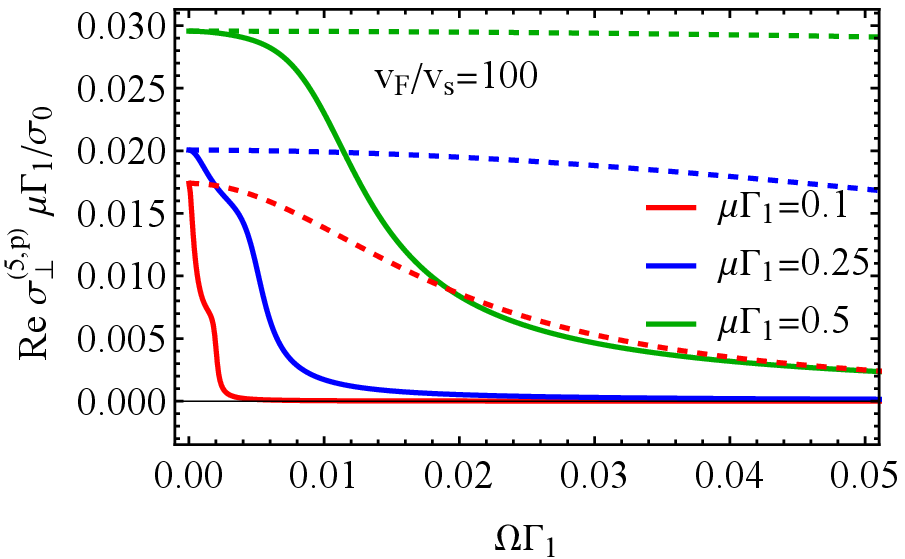}}
\hspace{0.01\textwidth}
\subfigure[]{\includegraphics[width=0.45\textwidth]{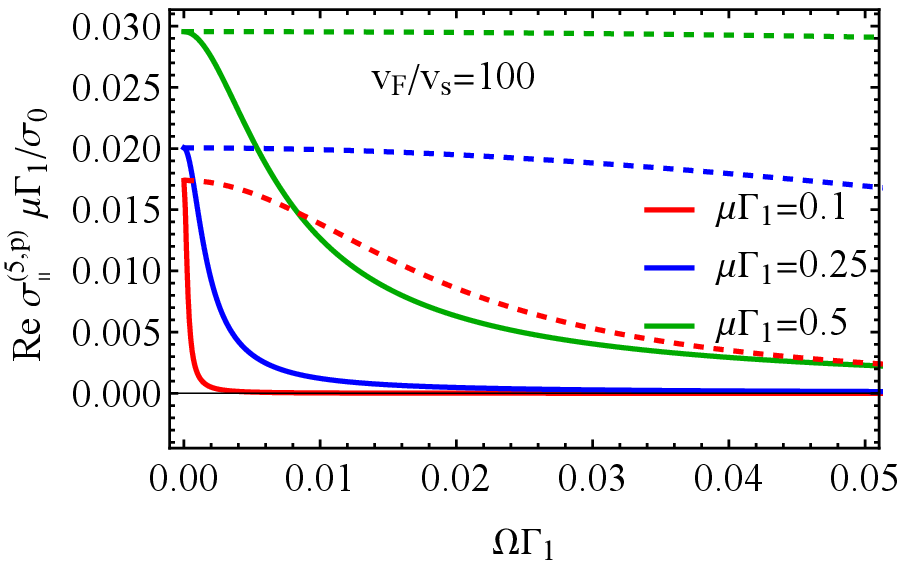}}
\vspace{-0.05cm}
\caption{
The dependence of the diagonal components of the optical conductivity on $\Omega \Gamma_1$ for a few values of $\mu \Gamma_1$. Panels (a) and (b) show $\text{Re}\,\sigma_{\perp}^{\rm (5,p)}(\Omega, \mathbf{q})$ and $\text{Re}\,\sigma_{\parallel}^{\rm (5,p)}(\Omega, \mathbf{q})$, respectively. Solid lines correspond to the case $q=\Omega/(\hbar v_s)$ and the dashed lines show the results at $q=0$. We used $\sigma_0=e^2\mu/(\hbar^2 v_F)$, $v_F/v_s=100$, and the broadening function (\ref{chiral-G-Gamma-tot}) with $\Gamma_0=0$.
}
\label{fig:Conductivity_Gamma_nonzero_q_T=0_Gamma1}
\end{figure*}

The dependence of the diagonal components of the chiral pseudoelectric conductivity on $\Omega \Gamma_1$ for local disorder $\Gamma(\omega) = \Gamma_1\omega^2$ is shown in Fig.~\ref{fig:Conductivity_Gamma_nonzero_q_T=0_Gamma1}. As in the case of the phenomenological constant broadening $\Gamma_0$, the results for $q=\Omega/(\hbar v_s)$ and $q=0$ are strongly different. The step-like feature in $\text{Re}\,\sigma_{\perp}^{\rm (5,p)}(\Omega, \mathbf{q})$ is, however, less pronounced, cf. Figs.~\ref{fig:Conductivity_Gamma_nonzero_q_T=0_Gamma0}(a) and \ref{fig:Conductivity_Gamma_nonzero_q_T=0_Gamma1}(a). Therefore, this step-like feature might not be easily observed in dirty samples.

Thus, we conclude that the characteristic features of the chiral pseudoelectric conductivity tensor at nonzero wave vector, $\mathbf{q}\neq \mathbf{0}$, persist beyond the clean case considered in Sec.~\ref{sec:chiral-clean}. In particular, we find that the chiral pseudoelectric conductivity is suppressed with frequency for a sound-like dispersion relation $q=\Omega/(\hbar v_s)$. In addition to the suppression, the transverse component of the conductivity tensor show a characteristic step-like feature determined by $v_F/v_s$, see Eq.~(\ref{chiral-sigma-D-first}), which could tolerate disorder.

\section{Estimates}
\label{sec:Estimates}

In this Section, we provide numerical estimates for the predicted effects. For an order of magnitude estimate, we use the parameters corresponding to the Weyl semimetal TaAs~\cite{Arnold-Felser:2016b,Zhang-Hasan-TaAs:2016,Laliberte-Quilliam:2019},
\begin{equation}
\label{Estimates-TaAs}
v_{\rm F}\approx 3\times 10^7~\mbox{cm/s}, \quad  \mu\approx 20~\mbox{meV}, \quad v_s \approx 2.8\times 10^5~\mbox{cm/s}.
\end{equation}
Notice that other materials with a simpler band structure might be used. For example, we could mention the Weyl semimetal EuCd$_2$As$_2$~\cite{Wang-Canfield:2019,Soh-Boothroyd:2019,Ma-Shi:2019}, which has a broken time-reversal symmetry and only two Weyl nodes.

We use the sound dispersion relation $q=\Omega/(\hbar v_s)$ and estimate the Weyl node separation as $b\approx 0.06\,\pi/a \approx 5.5\times10^{6}~\mbox{cm}^{-1}$~\cite{Lv-Ding-TaAs:2015b} where $a\approx 3.435~\text{\AA}$ is the lattice constant of TaAs. The separation is expected to be of the same order or even larger for magnetic Weyl semimetals. For sound-induced axial gauge fields, the wave vector can be estimated as $q/\left(1~\mbox{cm}^{-1}\right)\approx 2.2\times10^3\left[\Omega/(2\pi\hbar)\right]/\left(1~\mbox{GHz}\right)$. Then, we obtain
\begin{equation}
\label{estimate-sigmaxy}
\frac{\sigma_{xy}^{\rm (APHE)}}{\sigma_{xy}^{\rm (AHE)}} \sim \frac{q}{b} \sim 4\times 10^{-3}.
\end{equation}
Therefore, we expect the APHE to be a few orders of magnitude weaker than the AHE in Weyl semimetals if electric and pseudoelectric fields have the same magnitude. On the other hand, since the typical sound frequencies are much lower than the Fermi energy $\mu \approx 20~\mbox{meV} \approx 4.8\times \hbar~\mbox{THz}$, the ``matter" contribution to the APHE given in Eq.~(\ref{el-sigma-xy-mu-neq-0-Main}) will provide the main contribution to the APHE and could enhance the corresponding conductivity.

Furthermore, since $E_5 \sim q\Omega$, the APHE has a profoundly different, cubic at $\Omega\gg\mu$, scaling with the frequency of external perturbations. This feature might be used to distinguish it from the AHE and other Hall-like responses. Furthermore, the direction of the APHE current is determined by the wave vector $\mathbf{q}$, which allows for a more complicated dependence on the type of the perturbations, e.g., longitudinal or transverse sound.

\begin{figure}[t]
\centering
\includegraphics[width=0.45\textwidth]{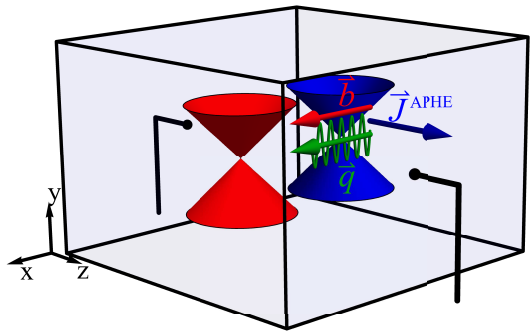}
\vspace{-0.05cm}
\caption{The model setup for measuring the APHE. The transverse deformations with the wave vector $\mathbf{q}$ induce the APHE electric current $\mathbf{J}^{\text{APHE}}\perp \mathbf{q}$.
}
\label{fig:estimate-model}
\end{figure}

Let us consider an explicit model for $\mathbf{A}_5$. According to Ref.~\cite{Cortijo-Vozmediano:2015}, the axial gauge field is related to the strain tensor and the chiral shift as $A_{5,j} = -c\hbar \beta u_{jl}b_l/e$~\footnote{For simplicity, we neglect the contribution of the deformation potential in $\mathbf{A}_5$.}, where $u_{jl}=i\left(q_ju_l+q_lu_j\right)/2$ is the symmetrized strain tensor and $\beta$ is the Gr\"{u}neisen parameter. Then, by using the APHE conductivity (\ref{el-sigma-xy-mu=0-fin}) and the expression for the pseudoelectric field $\mathbf{E}_5(\Omega,\mathbf{q}) = i\Omega \mathbf{A}_5(\Omega,\mathbf{q})/(\hbar c)$, we obtain the following APHE current:
\begin{equation}
\label{estimate-J-APHE}
\mathbf{J}^{\text{APHE}} = \frac{e}{8\pi^2 \hbar} \beta \Omega \left[\mathbf{u}\times \mathbf{q}\right] \left(\mathbf{q}\cdot \mathbf{b}\right) \approx 12\times \left(\frac{\Omega/(2\pi \hbar)}{1~\mbox{GHz}}\right)^3~\frac{\mu\mbox{A}}{\text{cm}^2},
\end{equation}
where we used $\beta\approx1$. As one can see, the APHE current reaches potentially measurable amplitudes and requires transverse deformations in the case under consideration. We present the model setup for measuring the APHE in Fig.~\ref{fig:estimate-model}.

In addition to semimetals, the APHE might be realized in metamaterials, where the speed of sound and the Fermi velocity are different from those in solids. The tunability of the effective Fermi velocity might allow one to realize unconventional chiral responses where the interband transitions can play a prominent role.

Finally, let us briefly discuss the chiral pseudoelectric response. The chiral current response is drastically different from its electric current counterpart: the corresponding chiral currents are manifested as a spin polarization. Such currents might be detected in an electric response by applying an external magnetic field and activating the chiral anomaly. Indeed, due to the continuity relation, nonuniform chiral current leads to time-dependent chiral charge density. In the presence of a magnetic field, the latter allows for an oscillating electric current by the chiral magnetic effect~\cite{Vilenkin:1980,Fukushima:2008} that could be measured via conventional electric current measurement.

\section{Summary and discussion}
\label{sec:Summary}

In this work, we investigated the electric and chiral current response to external time and coordinate dependent pseudoelectric field. The latter can be induced by dynamical perturbations such as sound or magnetic waves. One of the key features that make our findings distinct from the conventional response to electromagnetic field is the dependence of the conductivity on the wave vector $\mathbf{q}$.

We found that the pseudoelectric field $\mathbf{E}_5$ induces a Hall-like current. In this case, the electric current, the pseudoelectric field, and the wave vector are mutually orthogonal. Since the pseudoelectric field originates from the dynamical separation between the Weyl nodes and the corresponding conductivity resembles that of the AHE with the Weyl node separation replaced by $\mathbf{q}$, we dubbed this Hall-like response the anomalous pseudo-Hall effect. While its magnitude is estimated to be smaller compared to that of the AHE in Weyl semimetals, the APHE has a distinctive cubic scaling with frequency of sound which can be used to experimentally pinpoint the effect.

Further, we studied a chiral or valley response to the pseudoelectric field. In the absence of chirality mixing terms, there is a similarity between the chiral response to the pseudoelectric field and an electric response to an electric field. However, the wave vector cannot be discarded in the former. Assuming a clean limit and vanishing temperature, we obtained a full chiral conductivity tensor focusing on the dependence on the wave vector; see Eqs.~(\ref{chiral-sigma-xx-IB})--(\ref{chiral-sigma-zz-D}). We showed that the intraband part of the conductivity is strongly modified by the wave vector where the Drude peak acquires a step-like feature and a finite width even in the clean limit; see Fig.~\ref{fig:Conductivity_intra_nonzero_q_T=0}. The interband part of the conductivity can be also nontrivial where the wave vector transforms a single step-like feature at $\Omega=2\mu$ into a region with a different slope. Contrary to the intraband part of the conductivity, the interband one, however, requires for its observation a rather exotic parameter regime with $v_F<v_s$. Therefore, the intraband processes are expected to be dominant in Weyl semimetals.

Finally, let us discuss a few simplifications employed in this study and future directions. In order to simplify the calculations and present the results in a concise form, we used a linearized model of a Weyl semimetal. The extension to the case of more realistic periodic models will require a more careful definition of axial fields and chiral currents. Further, we considered the limit of vanishing temperature. We expect, however, that the effect of temperature will be qualitatively unimportant and will lead to a broadening of step-like features in the conductivity tensor components.

\begin{acknowledgments}
The authors acknowledge useful communications and discussions with I.A.~Shovkovy and F.~Pe\~{n}a-Benitez. P.O.S. acknowledges support through the Yale Prize Postdoctoral Fellowship in Condensed Matter Theory. The work of E.V.G. was partially supported by the Program of Fundamental Research of the Physics and Astronomy Division of the National Academy of Sciences of Ukraine.
\end{acknowledgments}

\onecolumngrid

\appendix

\section{Spectral function}
\label{sec:app-1}

In this appendix, we present the Green and spectral functions for the model defined in Sec.~\ref{sec:electric-Model}. The retarded ($+$) and advanced ($-$) Green function read as
\begin{equation}
\label{app-GRA-def}
\hat{G}\left(\omega\pm i0,\mathbf{k}\right) = \frac{i}{\left(\omega +\mu \pm i0\right)\hat{I} -\hat{H} }.
\end{equation}
Here, $\hat{H}=\text{diag}\left(\hat{H}_{+},\hat{H}_{-} \right),\,\,\hat{H}_{\lambda=\pm}=\lambda \hbar v_F \bm{\sigma} \cdot \left(\mathbf{k}-\lambda \mathbf{b} \right)$, $\omega$ is the frequency (measured in energy units), $\mu$ is the chemical potential, $\lambda$ is the chirality of the Weyl node, $v_F$ is the Fermi velocity, $\mathbf{k}$ is momentum, $\mathbf{b}$ defines the separation between the Weyl nodes (also known as the chiral shift), and $\bm{\sigma}$ is the vector of the Pauli matrices acting in the pseudospin space.

The spectral function is defined as the difference between the advanced and retarded Green functions
\begin{equation}
\label{app-A-def}
\hat{A}(\omega; \mathbf{k}) =\frac{1}{2\pi} \left[\hat{G}(\omega +i0; \mathbf{k}) -\hat{G}(\omega -i0; \mathbf{k}) \right]_{\mu=0} = \sum_{\lambda=\pm} \hat{A}_{\lambda}(\omega;\mathbf{k}) \frac{\hat{1} +\lambda \gamma^5}{2},
\end{equation}
where $\gamma^5=\text{diag}{\left(\hat{I}, -\hat{I}\right)}$ is the chirality matrix. The explicit form of $\hat{A}_{\lambda}(\omega;\mathbf{k})$ is given in Eq.~(\ref{model-spectral-function-def}), i.e.,
\begin{equation}
\label{app-spectral-function-def}
\hat{A}_{\lambda}(\omega;\mathbf{k}) = \frac{1}{2}\left[\delta(\omega - \varepsilon_{\mathbf{k},\lambda}) +\delta(\omega + \varepsilon_{\mathbf{k},\lambda})\right] \hat{I} +\frac{\lambda \hbar v_{F}}{2\varepsilon_{\mathbf{k},\lambda}} \left[\bm{\sigma} \cdot \left(\mathbf{k}-\lambda \mathbf{b}\right)\right] \left[\delta(\omega - \varepsilon_{\mathbf{k},\lambda}) -\delta(\omega + \varepsilon_{\mathbf{k},\lambda})\right].
\end{equation}
Here, $\varepsilon_{\mathbf{k},\lambda}=\hbar v_F |\mathbf{k}- \lambda \mathbf{b}|$ is the dispersion relation of quasiparticles with chirality $\lambda$.

The effects of disorder can be taken into account via the self-energy corrections in the Green functions (see, e.g., Refs.~\cite{Ashby-Carbotte:2014-opt,Tabert-Carbotte:2016}). This is equivalent to replacing $i0\to i\Gamma(\omega)$ in Eq.~(\ref{app-GRA-def}) where $\Gamma(\omega)$ is the broadening function whose dependence on $\omega$ is determined by the type of scatterers. The simplest model is to take constant broadening $\Gamma(\omega)=\Gamma_0$. A more realistic model with local scatterers leads to the frequency-dependent broadening  $\Gamma(\omega) = \Gamma_1 \omega^2$~\cite{Burkov-Hook:2011}.

Nonvanishing broadening can be accounted for by replacing the $\delta$-functions in the spectral function (\ref{app-spectral-function-def}) with the Lorentzian function, i.e.,
\begin{equation}
\label{app-Lorentzian}
\delta\left(\omega -\varepsilon_{\mathbf{k},\lambda}\right) \to \delta_{\Gamma}\left(\omega -\varepsilon_{\mathbf{k},\lambda}\right) = \frac{1}{\pi}\frac{\Gamma(\omega)}{\left(\omega -\varepsilon_{\mathbf{k},\lambda}\right)^2+\Gamma^2(\omega)}.
\end{equation}

\section{Electric conductivity tensor}
\label{sec:app-2}

In this appendix, we present the details of the calculation of the pseudoelectric conductivity tensor discussed in Sec.~\ref{sec:electric-sigma}. The real part of the conductivity is defined in Eq.~(\ref{el-sigma-ij-def}). In the integrand of $\sigma_{ij}^{\rm (p)}(\Omega,\mathbf{q})$, we have
\begin{eqnarray}
&&\text{Tr} \left[\hat{v}_i \hat{A}(\mathbf{k},\omega) \gamma^5 \hat{v}_{j}  \hat{A}(\mathbf{k}+\mathbf{q},\omega^{\prime})\right]
= \sum_{\lambda=\pm} \frac{\lambda v_{F}^2}{2} \Bigg\{ \Bigg[ \frac{(k_i-\lambda b_i)(k_j+q_j-\lambda b_j)+(k_i+q_i-\lambda b_i)(k_j-\lambda b_j)}{|\mathbf{k}-\lambda \mathbf{b}||\mathbf{k}+\mathbf{q}-\lambda \mathbf{b}|} \nonumber\\
&&-\delta_{ij}\frac{\left(\mathbf{k}-\lambda \mathbf{b}\right) \cdot \left(\mathbf{k}+\mathbf{q}-\lambda \mathbf{b}\right)}{|\mathbf{k}-\lambda \mathbf{b}||\mathbf{k}+\mathbf{q}-\lambda \mathbf{b}|}\Bigg][\delta (\omega - \varepsilon_{\mathbf{k},\lambda}) -\delta (\omega + \varepsilon_{\mathbf{k},\lambda})][\delta (\omega' - \varepsilon_{\mathbf{k}+\mathbf{q},\lambda}) -\delta (\omega^{\prime}+ \varepsilon_{\mathbf{k}+\mathbf{q},\lambda})] \nonumber\\
&&+\delta_{ij} \left[\delta(\omega - \varepsilon_{\mathbf{k},\lambda}) +\delta (\omega + \varepsilon_{\mathbf{k},\lambda})\right]\left[\delta (\omega^{\prime} - \varepsilon_{\mathbf{k}+\mathbf{q},\lambda}) +\delta (\omega^{\prime}+ \varepsilon_{\mathbf{k}+\mathbf{q},\lambda})\right] \nonumber\\
&&- i \lambda \varepsilon_{ijl} \frac{k_l-\lambda b_l}{|\mathbf{k}-\lambda \mathbf{b}|} [\delta (\omega- \varepsilon_{\mathbf{k},\lambda})- \delta (\omega+ \varepsilon_{\mathbf{k},\lambda})] [\delta (\omega' - \varepsilon_{\mathbf{k}+\mathbf{q},\lambda}) +\delta (\omega'+ \varepsilon_{\mathbf{k}+\mathbf{q},\lambda})] \nonumber\\
&&+ i \lambda \varepsilon_{ijl} \frac{k_l+q_l-\lambda b_l}{|\mathbf{k}+\mathbf{q}-\lambda \mathbf{b}|} [\delta (\omega - \varepsilon_{\mathbf{k},\lambda}) +\delta (\omega+ \varepsilon_{\mathbf{k},\lambda})][\delta (\omega'- \varepsilon_{\mathbf{k}+\mathbf{q},\lambda})- \delta (\omega'+ \varepsilon_{\mathbf{k}+\mathbf{q},\lambda})] \Bigg\}.
\end{eqnarray}

For simplicity, in this appendix, we assume that $\mathbf{q}=(0,0,q_z)$. Let us start with the diagonal components of the conductivity tensor. We have the following interband and intraband parts:
\begin{eqnarray}
\text{Re}\,\sigma_{ii}^{\rm (p), \text{inter}}(\Omega,\mathbf{q}) &=& \frac{e^2 \hbar v_F^2}{16 \pi^2 \Omega} \sum_{\lambda=\pm} \lambda \int_{-\infty}^{+\infty}d\omega \left[f(\omega)-f(\omega+\Omega)\right] \nonumber\\
&\times&\int d^3 k
\left\{\frac{\left[(\mathbf{k}-\lambda \mathbf{b}) \cdot (\mathbf{k}+\mathbf{q}-\lambda \mathbf{b})\right]-2 (k_i-\lambda b_i)(k_i+q_i- \lambda b_i)}{|\mathbf{k}-\lambda\mathbf{b}||\mathbf{k}+\mathbf{q}-\lambda\mathbf{b}|} +1 \right\} \nonumber\\
&\times& \left[\delta(\omega - \varepsilon_{\mathbf{k},\lambda})\delta(\omega +\Omega + \varepsilon_{\mathbf{k}+\mathbf{q},\lambda})
+\delta(\omega + \varepsilon_{\mathbf{k},\lambda})\delta(\omega+\Omega - \varepsilon_{\mathbf{k}+\mathbf{q},\lambda})\right]
\end{eqnarray}
and
\begin{eqnarray}
\text{Re}\,\sigma_{ii}^{\rm (p), \text{intra}}(\Omega,\mathbf{q}) &=& \frac{e^2 \hbar v_F^2}{16 \pi^2 \Omega} \sum_{\lambda=\pm}\lambda \int_{-\infty}^{+\infty}d\omega \left[f(\omega)-f(\omega+\Omega)\right] \nonumber\\
&\times&\int d^3 k
\left\{\frac{2(k_i-\lambda b_i)(k_i+q_i- \lambda b_i) -\left[(\mathbf{k}-\lambda \mathbf{b}) \cdot (\mathbf{k}+\mathbf{q}-\lambda \mathbf{b})\right]}{|\mathbf{k}-\lambda\mathbf{b}||\mathbf{k}+\mathbf{q}-\lambda\mathbf{b}|} +1 \right\} \nonumber\\
&\times& \left[\delta(\omega - \varepsilon_{\mathbf{k},\lambda})\delta(\omega +\Omega -\varepsilon_{\mathbf{k}+\mathbf{q},\lambda})
+\delta(\omega + \varepsilon_{\mathbf{k},\lambda})\delta(\omega+\Omega + \varepsilon_{\mathbf{k}+\mathbf{q},\lambda})\right].
\end{eqnarray}

Since we have a preferred direction determined by $\mathbf{q}$, it is convenient to use cylindrical coordinates with $k_x=k_{\perp} \cos{\varphi}$ and $k_y=k_{\perp} \sin{\varphi}$.
Integrating over $\varphi$, $k_{\perp}$, and $k_{z}$, we obtain that all diagonal components of the pseudoelectric conductivity tensor vanish after summation over Weyl nodes.

Let us proceed to the off-diagonal components of the conductivity tensor. In the limit $T \rightarrow 0$, we obtain
\begin{eqnarray}
\label{el-sigma-xy-mu}
\sigma_{ij}^{\rm (p)}(\Omega,\mathbf{q}) &=& \varepsilon_{ijl} \frac{\hbar e^2 v_F^2 }{\Omega} \sum_{\lambda=\pm}\text{v.p.} \int \frac{d^3 k}{(2\pi)^3} \Bigg\{ \frac{\Omega}{\left(\varepsilon_{\mathbf{k},\lambda}+\varepsilon_{\mathbf{k}+\mathbf{q},\lambda} \right)^2-\Omega^2}\left( \frac{k_l+q_l -\lambda b_l}{|\mathbf{k}+\mathbf{q} -\lambda\mathbf{b}|}+ \frac{k_l -\lambda b_l}{|\mathbf{k} -\lambda\mathbf{b}|}\right) \nonumber \\
&+& \left[ \frac{k_l+q_l -\lambda b_l}{|\mathbf{k}+\mathbf{q} -\lambda\mathbf{b}|}\frac{\varepsilon_{\mathbf{k}+\mathbf{q},\lambda}}{\varepsilon_{\mathbf{k}+\mathbf{q},\lambda}^2-\left(\varepsilon_{\mathbf{k},\lambda}+\Omega \right)^2}-\frac{k_l -\lambda b_l}{|\mathbf{k} -\lambda\mathbf{b}|}\frac{\varepsilon_{\mathbf{k},\lambda}+\Omega}{\varepsilon_{\mathbf{k}+\mathbf{q},\lambda}^2-\left(\varepsilon_{\mathbf{k},\lambda}+\Omega \right)^2}\right]\Theta(\mu-\varepsilon_{\mathbf{k},\lambda}) \nonumber\\
&+& \left[ \frac{k_l+q_l -\lambda b_l}{|\mathbf{k}+\mathbf{q} -\lambda\mathbf{b}|}\frac{\varepsilon_{\mathbf{k}+\mathbf{q},\lambda}-\Omega}{\varepsilon_{\mathbf{k},\lambda}^2-\left(\varepsilon_{\mathbf{k}+\mathbf{q},\lambda}-\Omega \right)^2}-\frac{k_l-\lambda b_l}{|\mathbf{k} -\lambda\mathbf{b}|}\frac{\varepsilon_{\mathbf{k},\lambda}}{\varepsilon_{\mathbf{k},\lambda}^2-\left(\varepsilon_{\mathbf{k}+\mathbf{q},\lambda}-\Omega \right)^2} \right]\Theta(\mu-\varepsilon_{\mathbf{k}+\mathbf{q},\lambda}) \Bigg\}.
\end{eqnarray}

For $\mu=0$, we derive
\begin{eqnarray}
\label{el-sigma-xy-mu=0}
\sigma_{xy}^{\rm (p)}(\Omega,\mathbf{q})&=& \hbar e^2 v_F^2  \sum_{\lambda=\pm}\text{v.p.} \int \frac{d^3 k}{(2 \pi)^3} \left( \frac{k_z+q_z -\lambda b}{|\mathbf{k}+\mathbf{q} -\lambda\mathbf{b}|} + \frac{k_z -\lambda b}{|\mathbf{k} -\lambda\mathbf{b}|} \right) \frac{1}{\left(\varepsilon_{\mathbf{k},\lambda}+\varepsilon_{\mathbf{k}+\mathbf{q},\lambda}\right)^2-\Omega^2} \nonumber\\
&=&\frac{e^2}{16\pi^2 \hbar \tilde{\Omega}^3} \sum_{\lambda=\pm} \text{v.p.} \int_{-\infty}^{+\infty} dk_z \Bigg\{ 2\tilde{\Omega} q_z \left(|k_z+q_z -\lambda b|-|k_z -\lambda b|\right) +\left(2k_z+q_z-2\lambda b\right) \left(q_z^2-\tilde{\Omega}^2\right) \nonumber\\
&\times& \left[ \tanh^{-1}\left(\frac{2 \tilde{\Omega} |k_z- \lambda b|}{(k_z+q_z -\lambda b)^2 -(k_z -\lambda b)^2-\tilde{\Omega}^2}\right)- \tanh ^{-1}\left(\frac{2 \tilde{\Omega} |k_z+q_z-b|}{(k_z+q_z- \lambda b)^2-(k_z- \lambda b)^2+\tilde{\Omega}^2}\right) \right] \Bigg\}\nonumber\\
&=& \frac{e^2}{ 8 \pi^2 \hbar} \sum_{\lambda=\pm}  \left(-2\lambda b+q_z\right)=\frac{e^2 q_z}{4 \pi^2 \hbar},
\end{eqnarray}
where $\tilde{\Omega}=\Omega/(\hbar v_F)$. In the case of an arbitrary orientation of $\mathbf{q}$, we have $\sigma_{ij}^{\rm (p)}(\Omega,\mathbf{q})=\varepsilon_{ijl} e^2 q_l/(4 \pi^2 \hbar)$.

Finally, the general expression for $\mu\neq0$ reads
\begin{eqnarray}
\label{el-sigma-xy-mu-neq-0}
\sigma_{xy}^{\rm (p)}(\Omega,\mathbf{q}) &=& \frac{e^2 q_z}{4 \pi^2 \hbar} + \frac{\hbar e^2 v_F^2 }{\Omega} \sum_{\lambda=\pm}\text{v.p.} \int \frac{d^3 k}{(2\pi)^3} \Bigg\{ \left[ \frac{k_z+q_z -\lambda b}{|\mathbf{k}+\mathbf{q} -\lambda\mathbf{b}|}\frac{\varepsilon_{\mathbf{k}+\mathbf{q},\lambda}}{\varepsilon_{\mathbf{k}+\mathbf{q},\lambda}^2-\left(\varepsilon_{\mathbf{k},\lambda}+\Omega \right)^2}-\frac{k_z -\lambda b}{|\mathbf{k} -\lambda\mathbf{b}|}\frac{\varepsilon_{\mathbf{k},\lambda}+\Omega}{\varepsilon_{\mathbf{k}+\mathbf{q},\lambda}^2-\left(\varepsilon_{\mathbf{k},\lambda}+\Omega \right)^2}\right] \nonumber\\
&\times&\Theta(\mu-\varepsilon_{\mathbf{k},\lambda}) + \left[ \frac{k_z+q_z -\lambda b}{|\mathbf{k}+\mathbf{q} -\lambda\mathbf{b}|}\frac{\varepsilon_{\mathbf{k}+\mathbf{q},\lambda}-\Omega}{\varepsilon_{\mathbf{k},\lambda}^2-\left(\varepsilon_{\mathbf{k}+\mathbf{q},\lambda}-\Omega \right)^2}-\frac{k_z-\lambda b}{|\mathbf{k} -\lambda\mathbf{b}|}\frac{\varepsilon_{\mathbf{k},\lambda}}{\varepsilon_{\mathbf{k},\lambda}^2-\left(\varepsilon_{\mathbf{k}+\mathbf{q},\lambda}-\Omega \right)^2} \right]\Theta(\mu-\varepsilon_{\mathbf{k}+\mathbf{q},\lambda}) \Bigg\}  \nonumber \\
&=&  \frac{e^2 q_z}{4 \pi^2 \hbar} \Bigg \{1-\frac{\Omega \left(\tilde{q}^2-\Omega^2\right)}{8  \tilde{q} ^3} \Bigg[ 4 \mu \Omega  \ln \left|\frac{\tilde{q}^2}{(\tilde{q}-\Omega)^2} \frac{ (\tilde{q} - \Omega +2  \mu ) (\tilde{q} - \Omega -2  \mu )}{ (\tilde{q} + \Omega -2  \mu) ( \tilde{q} + \Omega + 2  \mu)}\right| -8 \mu \tilde{q}+ \left(\tilde{q}^2-\Omega^2 -4 \mu^2 \right) \nonumber\\
&\times&\ln \left| \frac{(\tilde{q} + \Omega +2 \mu) (\tilde{q} - \Omega +2 \mu)}{(\tilde{q} + \Omega-2 \mu)(\tilde{q} - \Omega-2  \mu)} \right| \Bigg] \Bigg\}.
\end{eqnarray}
Here, $\tilde{q}=\hbar v_F q$. By using the sound dispersion relation $\Omega=\hbar v_s q$, Eq.~(\ref{el-sigma-xy-mu-neq-0}) can be rewritten as follows:
\begin{eqnarray}
\label{el-sigma-xy-mu-neq-0-1}
\sigma_{xy}^{\rm (p)}(\Omega,\mathbf{q}) &=& \frac{e^2 q_z}{4 \pi^2 \hbar} \Bigg \{1-\frac{\left(v_F^2-v_s^2\right)}{8 v_F^3 v_s^3 \Omega ^2} \Bigg[ 4 \mu \Omega  v_s^4 \ln \left|\frac{v_F^2}{(v_F-v_s)^2} \frac{ (v_F \Omega -v_s \Omega +2 v_s \mu ) (v_F \Omega -v_s \Omega -2 v_s \mu )}{ (v_F \Omega +v_s \Omega -2 v_s \mu) ( v_F \Omega +v_s \Omega + 2 v_s \mu)}\right| \nonumber\\
&-&8 \mu \Omega  v_s^3 v_F+v_s^2 \left(\Omega^2 \left(v_F^2-v_s^2\right)-4 \mu^2 v_s^2\right) \ln \left| \frac{(v_F \Omega +v_s \Omega +2 v_s \mu) (v_F \Omega -v_s \Omega+2 v_s \mu)}{(v_F \Omega +v_s \Omega-2 v_s \mu)(v_F \Omega -v_s \Omega-2 v_s \mu)}
\right| \Bigg] \Bigg\} \nonumber \\
&=&\frac{e^2 q_z}{4 \pi^2 \hbar} \left[1+ \frac{\mu}{\Omega}\left(1+\frac{8}{3} \frac{\mu^2}{\Omega^2} \right)\frac{v_s^2}{v_F^2} +o \left(\frac{v_s^2}{v_F^2}\right)  \right],
\end{eqnarray}
where we expanded in small $v_s/v_F\ll1$.

\section{Chiral conductivity tensor}
\label{sec:app-3}

\subsection{Real part}
\label{sec:app-3-Re}

As we discussed at the beginning of Sec.~\ref{sec:chiral}, the chiral pseudoelectric conductivity tensor $\sigma_{ij}^{(5,p)}(\Omega,\mathbf{q})$ can be straightforwardly obtained from Eqs.~(\ref{model-sigma-ij-def}) and (\ref{model-Pi-ij-def}) by replacing $\hat{v}_i \to \gamma^5 \hat{v}_i$ in Eq.~(\ref{model-Pi-ij-def}).

Let us start with the real part of $\sigma_{ij}^{(5,p)}(\Omega,\mathbf{q})$. We obtain the following expression:
\begin{eqnarray}
\text{Re}\, \sigma_{ij}^{(5,p)}(\Omega,\mathbf{q}) &=& \frac{\hbar e^2}{\Omega} \text{v.p.} \int_{-\infty}^{+\infty}d\omega^{\prime} \int_{-\infty}^{+\infty} d\omega \frac{f(\omega)-f(\omega^{\prime})}{\omega^{\prime}-\omega-\Omega} \int \frac {d^3 k}{(2\pi)^3} \mbox{Im} \left\{ \text{Tr}\left[\hat{v}_i \hat{A}(\mathbf{k},\omega) \hat{v}_j \hat{A}(\mathbf{k}+\mathbf{q},\omega^{\prime})\right] \right\} \nonumber\\
&+&\frac{\pi \hbar e^2}{\Omega} \int_{-\infty}^{+\infty}d\omega \left[f(\omega)-f(\omega+\Omega)\right] \int \frac{d^3 k}{(2\pi)^3} \mbox{Re} \left\{ \text{Tr}\left[\hat{v}_i \hat{A}(\mathbf{k},\omega) \hat{v}_j  \hat{A}(\mathbf{k}+\mathbf{q},\omega+\Omega)\right] \right\}.
\end{eqnarray}
Here, $f(\omega)=1/\left[e^{\left(\omega - \mu\right)/T}+1\right]$ is the Fermi-Dirac distribution function, $T$ is temperature, $\hat{v}_i = \partial_{k_i} \hat{H}/\hbar$ is the velocity operator, and $\hat{A}$ is the spectral function given in Eq.~(\ref{app-spectral-function-def}). In what follows, we mostly focus on the case of temperature small compared to the Fermi energy, i.e., we set $T\to0$. Finally, we include a finite wave vector of external perturbation $\mathbf{q}$. For the sake of definiteness, we assume that $\mathbf{q}=(0,0,q_z)$.

We find it convenient to separate the intraband and interband contributions to the conductivity tensor. For this, we use the explicit form of the spectral function given in Eq.~(\ref{app-spectral-function-def}) and separate the terms at $\delta(\omega - \varepsilon_{\mathbf{k},\lambda})\delta(\omega+\Omega - \varepsilon_{\mathbf{k}+\mathbf{q},\lambda})$ and $\delta(\omega - \varepsilon_{\mathbf{k},\lambda})\delta(\omega+\Omega + \varepsilon_{\mathbf{k}+\mathbf{q},\lambda})$. The interband and intraband parts of the conductivity tensor are given by
\begin{eqnarray}
\text{Re}\,\sigma_{ii}^{(5,p), \text{inter}}(\Omega,\mathbf{q}) &=& \frac{e^2 \hbar v_F^2}{16 \pi^2 \Omega} \sum_{\lambda=\pm}\int_{-\infty}^{+\infty}d\omega \left[f(\omega)-f(\omega+\Omega)\right] \int d^3 k \nonumber\\
&\times&\left\{\frac{\left[(\mathbf{k}-\lambda \mathbf{b}) \cdot (\mathbf{k}+\mathbf{q}-\lambda \mathbf{b})\right]-2 (k_i-\lambda b_i)(k_i+q_i- \lambda b_i)}{|\mathbf{k}-\lambda\mathbf{b}||\mathbf{k}+\mathbf{q}-\lambda\mathbf{b}|} +1 \right\} \nonumber\\
&\times& \left[\delta(\omega - \varepsilon_{\mathbf{k},\lambda})\delta(\omega +\Omega + \varepsilon_{\mathbf{k}+\mathbf{q},\lambda})
+\delta(\omega + \varepsilon_{\mathbf{k},\lambda})\delta(\omega+\Omega - \varepsilon_{\mathbf{k}+\mathbf{q},\lambda})\right]
\end{eqnarray}
and
\begin{eqnarray}
\text{Re}\,\sigma_{ii}^{(5,p), \text{intra}}(\Omega,\mathbf{q}) &=& \frac{e^2 \hbar v_F^2}{16 \pi^2 \Omega} \sum_{\lambda=\pm}\int_{-\infty}^{+\infty}d\omega \left[f(\omega)-f(\omega+\Omega)\right] \int d^3 k \nonumber\\
&\times&\left\{\frac{2(k_i-\lambda b_i)(k_i+q_i- \lambda b_i) -\left[(\mathbf{k}-\lambda \mathbf{b}) \cdot (\mathbf{k}+\mathbf{q}-\lambda \mathbf{b})\right]}{|\mathbf{k}-\lambda\mathbf{b}||\mathbf{k}+\mathbf{q}-\lambda\mathbf{b}|} +1 \right\} \nonumber\\
&\times& \left[\delta(\omega - \varepsilon_{\mathbf{k},\lambda})\delta(\omega +\Omega -\varepsilon_{\mathbf{k}+\mathbf{q},\lambda})
+\delta(\omega + \varepsilon_{\mathbf{k},\lambda})\delta(\omega+\Omega + \varepsilon_{\mathbf{k}+\mathbf{q},\lambda})\right].
\end{eqnarray}
Integrating over momenta and $\omega$, we derive the results given in Eqs.~(\ref{chiral-sigma-xx-IB})--(\ref{chiral-sigma-zz-D}) in the main text.

\begin{figure*}[ht]
\centering
\subfigure[]{\includegraphics[width=0.45\textwidth]{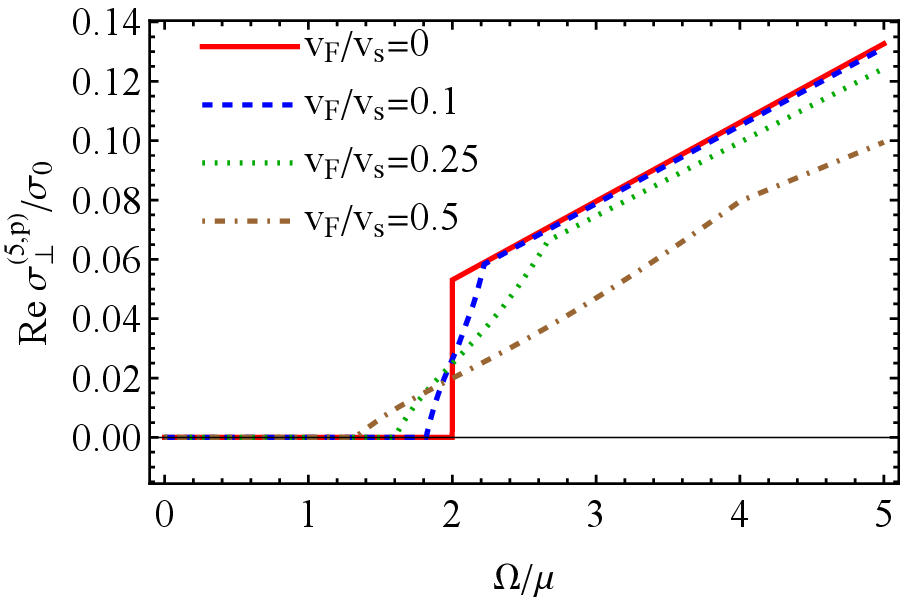}}
\hspace{0.01\textwidth}
\subfigure[]{\includegraphics[width=0.45\textwidth]{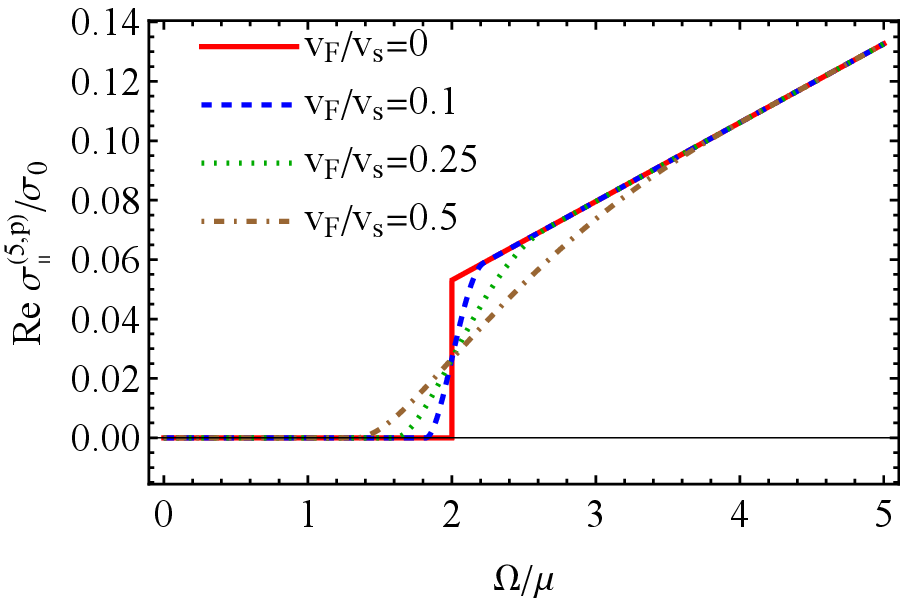}}
\vspace{-0.05cm}
\caption{
The dependence of the diagonal components of the chiral pseudoelectric conductivity on $\Omega/\mu$ for a few values of $v_F/v_s$. Panels (a) and (b) show $\text{Re}\,\sigma_{\perp}^{\rm (5,p)}(\Omega, \mathbf{q})$ and $\text{Re}\,\sigma_{\parallel}^{\rm (5,p)}(\Omega, \mathbf{q})$, respectively. We used $\sigma_0=e^2\mu/(\hbar^2 v_F)$. Notice that the intraband components of the conductivity tensor components vanish at $v_F/v_s<1$.
}
\label{fig:Conductivity_inter_nonzero_q_T=0}
\end{figure*}

The dependence of the diagonal interband components of the conductivity along the directions perpendicular and parallel to $\mathbf{q}$, i.e., $\sigma_{\perp}^{(5,p), \text{inter}}(\Omega,\mathbf{q}) \equiv \sigma_{xx}^{(5,p), \text{inter}}(\Omega,\mathbf{q})$ and $\sigma_{\parallel}^{(5,p), \text{inter}}(\Omega,\mathbf{q}) \equiv \sigma_{zz}^{(5,p), \text{inter}}(\Omega,\mathbf{q})$, on $\Omega$ is shown in Fig.~\ref{fig:Conductivity_inter_nonzero_q_T=0} for a few values of $v_F/v_s$. Since typically $v_F/v_s>1$ in semimetals, the presented results correspond to metamaterials. Overall, a step-like dependence of the interband contribution to the chiral pseudoelectric conductivity and the linear scaling with frequency in Fig.~\ref{fig:Conductivity_inter_nonzero_q_T=0} are the same as in the optical conductivity of Weyl semimetals~\cite{Tabert-Carbotte:2016}. Physically, this step-like dependence is a manifestation of the Pauli blocking principle, which forbids transitions between fully-occupied states. As follows from Eqs.~(\ref{chiral-sigma-xx-IB}) and (\ref{chiral-sigma-zz-IB}), the step-like feature at $\Omega=2\mu$ is transformed into an interpolating region where $\sigma_{ii}^{(5,p),\text{inter}}$ scales differently with $\Omega$. This region is determined by
\begin{equation}
\label{chiral-sigma-IB-region}
\frac{v_s}{v_F}\leq \frac{\Omega}{\mu} \leq \frac{2v_F}{v_s -v_F}.
\end{equation}
For small deviations $|\delta \Omega|\ll1$ where $\delta \Omega =\Omega/\mu-2$, both $\sigma_{\perp}^{(5,p),\text{inter}}$ and $\sigma_{\parallel}^{(5,p),\text{inter}}$ scale linearly with $\delta \Omega$ with the slope determined by $v_F/v_s$. The appearance of the region with a different slope is similar to the effects of tilt of the Weyl nodes in the electric conductivity~\cite{Carbotte:2016-tilt}.

It is instructive to show that the conventional Drude peak is restored in Eq.~(\ref{chiral-sigma-xx-D}) at $q\to0$. Indeed, we have the following limit in the weak sense:
\begin{equation}
\label{app-3-limit}
\frac{1}{\tilde{q}}\Theta (\tilde{q}-\Omega) \xrightarrow[\tilde{q} \rightarrow 0]{w} \delta (\Omega).
\end{equation}
It can be easily proven by integrating the left- and right-hand sides of the above equation. Performing the same integration in $\sigma_{\perp}^{\rm (5,p), \text{intra}}(\Omega,\mathbf{q}) \equiv \sigma_{xx}^{(5,p), \text{intra}}(\Omega,\mathbf{q})$, we obtain
\begin{equation}
\label{app-3-limit-sigma}
\lim_{\tilde{q} \to 0} \int_{-\infty}^{+\infty} d\Omega\, \text{Re}\, \sigma_{\perp}^{(5,p), \text{intra}}(\Omega,\mathbf{q}) =\lim_{\tilde{q} \rightarrow 0} \frac{e^2}{\pi \hbar^2 v_F}
\left(\frac{\mu^2}{3} +\frac{4}{45}\tilde{q}^2\right)= \frac{e^2 \mu^2}{3 \pi \hbar^2 v_F} =\int_{-\infty}^{+\infty} d\Omega \frac{e^2 \mu^2}{3 \pi \hbar^2 v_F} \delta(\Omega).
\end{equation}
Therefore,
\begin{equation}
\label{app-3-limit-1}
\text{Re}\, \sigma_{\perp}^{(5,p), \text{intra}}(\Omega,\mathbf{q}) \xrightarrow[\tilde{q} \rightarrow 0]{w} \frac{e^2 \mu^2}{3 \pi \hbar^2 v_F} \delta(\Omega).
\end{equation}

For the off-diagonal components at $\mu=0$, we obtain
\begin{eqnarray}
\text{Re}\, \sigma_{xy}^{(5,p)}(\Omega,\mathbf{q}) &=&
\hbar e^2 v_F^2  \sum_{\lambda=\pm} \lambda \, \text{v.p.} \int \frac {d^3 k}{(2 \pi)^3} \left( \frac{k_z+q_z-\lambda b}{|\mathbf{k}+\mathbf{q} -\lambda\mathbf{b}|} + \frac{k_z-\lambda b}{|\mathbf{k} -\lambda\mathbf{b}|} \right)\frac{1}{\left(\varepsilon_{\mathbf{k},\lambda}+\varepsilon_{\mathbf{k}+\mathbf{q},\lambda}\right)^2-\Omega^2} \nonumber\\
&=& \frac{e^2}{16 \pi^2 \hbar \tilde{\Omega}^3} \sum_{\lambda=\pm} \lambda \, \text{v.p.} \int_{-\infty}^{+\infty} d k_z \Bigg\{ 2 \tilde{\Omega} q_z \left(|k_z+q_z-\lambda b|-|k_z-\lambda b| \right) +(2k_z+q_z-2\lambda b) \left(q_z^2-\tilde{\Omega}^2\right) \nonumber\\
&\times& \left[\tanh^{-1}\left(\frac{2 \tilde{\Omega} |k_z- \lambda b|}{(k_z+q_z-\lambda b)^2-(k_z-\lambda b)^2-\tilde{\Omega}^2}\right)- \tanh^{-1}\left(\frac{2 \tilde{\Omega} |k_z+q_z-\lambda b|}{(k_z+q_z-\lambda b)^2-(k_z-\lambda b)^2+\tilde{\Omega}^2}\right) \right] \Bigg\}\nonumber\\
&=& \frac{e^2}{8 \pi^2 \hbar} \sum_{\lambda=\pm} \lambda \left(-2\lambda b+q_z\right)=-\frac{e^2 b}{2 \pi^2 \hbar},
\end{eqnarray}
where $\tilde{\Omega}=\Omega/(\hbar v_F)$. This chiral pseudoelectric conductivity has the same form as in the anomalous Hall effect~\cite{Yang-Lu:2011,Burkov-Balents:2011,Burkov-Hook:2011}.

\subsection{Imaginary part}
\label{sec:app-3-Im}

Finally, let us present the expressions for the imaginary part of the chiral pseudoelectric conductivity tensor. By using the Kramers-Kronig relation
\begin{equation}
\label{chiral-sigma-Kramers-Kronig}
\mbox{Im}\,\sigma_{ij}(\Omega,\mathbf{q}) = -\frac{2\Omega}{\pi}\, \mbox{v.p.} \int_{0}^{\Lambda} d\omega \frac{\mbox{Re}\,\sigma_{ij}(\omega,\mathbf{q})}{\omega^2-\Omega^2},
\end{equation}
where $\Lambda$ is a cutoff, and the expressions in Eqs.~(\ref{chiral-sigma-xx-IB})--(\ref{chiral-sigma-zz-D}), we obtain
\begin{eqnarray}
&&\text{Im}\, \sigma_{\perp}^{(5,p), \text{inter}}(\Omega,\mathbf{q}) = - \frac{e^2 }{12 \pi^2  \hbar^2 v_F \Omega } \left[\left(\Omega ^2-\tilde{q}^2\right) \ln \left|\frac{\Lambda ^2-\Omega ^2}{(2 \mu +\tilde{q})^2-\Omega ^2}\right|+2 \tilde{q}^2 \ln \left|\frac{\Lambda }{2 \mu +\tilde{q}}\right|\right] - \frac{e^2}{288 \pi^2  \hbar^2 \tilde{q}^3 v_F \Omega} \nonumber\\
&&\times \Bigg\{ 4 \tilde{q} \Omega ^2 \left[3 \left(4 \mu ^2+\Omega ^2\right)+7 \tilde{q}^2\right]\Theta (\mu-\tilde{q})+ 4 \mu  \Omega ^2 \left(22 \mu ^2+9 \tilde{q}^2-12 \mu  \tilde{q}+3 \Omega ^2\right)\Theta (\tilde{q}- \mu)+3 \Omega  \left(\tilde{q}^2-\Omega ^2\right) \nonumber\\
&&\times \left(12 \mu ^2+3 \tilde{q}^2+\Omega ^2\right)  \left[ \Theta ( \mu- \tilde{q}) \ln \left|\frac{(\Omega+2 \mu +\tilde{q} ) (\Omega-2 \mu +\tilde{q} )}{(\Omega -2 \mu -\tilde{q}) (\Omega+2 \mu -\tilde{q} )}\right|+\Theta (\tilde{q}-\mu) \ln \left|\frac{(\Omega-\tilde{q} ) (\Omega+2 \mu +\tilde{q} )}{(\Omega+\tilde{q} ) (\Omega-2 \mu -\tilde{q} )}\right|\right] \nonumber\\
&&- 6 \left(\tilde{q}^2-\Omega ^2\right) \left(4 \mu ^3+3 \mu  \Omega ^2-2 \tilde{q}^3+3 \mu  \tilde{q}^2\right) \left[ \Theta( \mu- \tilde{q}) \ln \left|\frac{(2 \mu -\tilde{q})^2-\Omega ^2}{(2 \mu +\tilde{q})^2-\Omega ^2}\right| + \Theta(\tilde{q}-\mu) \ln \left|\frac{\tilde{q}^2-\Omega ^2}{(2 \mu +\tilde{q})^2-\Omega ^2}\right| \right] \nonumber\\
&&+ 12 \left(2 \tilde{q}^5-3 \mu  \tilde{q}^4-4 \mu ^3 \tilde{q}^2\right)\left[ \Theta(\mu-\tilde{q}) \ln \left|\frac{2 \mu +\tilde{q}}{2 \mu -\tilde{q}}\right|  + \Theta (\tilde{q}- \mu) \ln \left|\frac{2 \mu +\tilde{q}}{\tilde{q}}\right| \right] \Bigg\},
\label{app-3-sigma-Im-xx-inter}
\end{eqnarray}

\begin{eqnarray}
&&\text{Im}\, \sigma_{\parallel}^{(5,p), \text{inter}}(\Omega,\mathbf{q}) =- \frac{e^2 \Omega}{12 \pi^2  \hbar^2 v_F  } \ln \left| \frac{\Lambda ^2-\Omega ^2}{\Omega^2-(2 \mu +\tilde{q})^2}\right| \nonumber\\
&&- \frac{e^2 \Omega}{144 \pi^2  \hbar^2 \tilde{q}^3 v_F} \Bigg\{ 4 \tilde{q} \left[8 \tilde{q}^2-3 \left(4 \mu ^2+\Omega^2\right)\right] \Theta ( \mu- \tilde{q})+4 \mu  \left(-22 \mu ^2+6 \tilde{q}^2+12 \mu  \tilde{q}-3 \Omega^2\right) \Theta(\tilde{q}-\mu) \nonumber\\
&&+3 \Omega \left(12 \mu ^2-3 \tilde{q}^2+\Omega^2\right) \left[ \Theta(\mu- \tilde{q}) \ln \left|\frac{(\Omega+2 \mu +\tilde{q}) (\Omega-2 \mu +\tilde{q})}{(\Omega-2 \mu -\tilde{q}) (\Omega+2 \mu -\tilde{q})}\right|+\Theta (\tilde{q}-\mu) \ln \left|\frac{(\Omega-\tilde{q}) (\Omega+2 \mu +\tilde{q})}{(\Omega+\tilde{q}) (\Omega-2 \mu -\tilde{q})}\right|\right] \nonumber\\
&&+6 \left(4 \mu ^3+\tilde{q}^3-3 \mu \tilde{q}^2+3 \mu  \Omega^2\right) \left[ \Theta(\mu- \tilde{q}) \ln \left|\frac{\Omega^2-(2 \mu +\tilde{q})^2}{\Omega^2-(2\mu-\tilde{q})^2}\right|  +\Theta (\tilde{q}- \mu)  \ln \left|\frac{\Omega^2-(2 \mu +\tilde{q})^2}{\Omega^2-\tilde{q}^2}\right| \right] \Bigg\},
\label{app-3-sigma-Im-zz-inter}
\end{eqnarray}

\begin{eqnarray}
&&\text{Im}\, \sigma_{\perp}^{(5,p), \text{intra}}(\Omega,\mathbf{q}) = \frac{e^2 \Theta (\mu-\tilde{q})}{144 \pi^2  \hbar^2 \tilde{q}^3 v_F }\left[2 \tilde{q} \Omega \left(36 \mu ^2+7 \tilde{q}^2+3 \Omega^2\right)+3 \left(\tilde{q}^2-\Omega^2\right) \left(12 \mu ^2+3 \tilde{q}^2+\Omega^2\right) \ln \left| \frac{\Omega+\tilde{q}}{\Omega-\tilde{q}}\right| \right] \nonumber\\
&&+ \frac{e^2 \Theta(\tilde{q}-\mu)\Theta (2\mu-\tilde{q})}{144 \pi^2  \hbar^2 \tilde{q}^3 v_F }  \left[2 \Omega (2 \mu -\tilde{q}) \left(40 \mu ^2+7 \tilde{q}^2-4 \mu  \tilde{q}+3 \Omega^2\right)+3 \left(\tilde{q}^2-\Omega^2\right) \left(12 \mu ^2+3 \tilde{q}^2+\Omega^2\right)  \ln \left|\frac{\Omega+2 \mu -\tilde{q}}{\Omega-2 \mu +\tilde{q}}\right|\right]  \nonumber\\
&&-\frac{e^2 \Theta (\tilde{q}-2\mu)}{288 \pi^2 \hbar^2 \tilde{q}^3 v_F\Omega}  \bigg[-4 \mu \Omega^2 \left(22 \mu ^2+9 \tilde{q}^2+12 \mu \tilde{q}+3 \Omega^2\right)+3 \Omega \left(\tilde{q}^2-\Omega^2\right) \left(12 \mu ^2+3 \tilde{q}^2+\Omega^2\right) \ln \left|\frac{(\Omega-\tilde{q}) (\Omega-2 \mu +\tilde{q})}{(\Omega+\tilde{q}) (\Omega+2 \mu -\tilde{q})}\right|\nonumber\\
&& +6 \left(\tilde{q}^2-\Omega^2\right) \left(-4 \mu ^3+2 \tilde{q}^3-3 \mu  \tilde{q}^2-3 \mu  \Omega^2\right) \ln \left|\frac{(\tilde{q}-2 \mu )^2-\Omega^2}{\tilde{q}^2-\Omega^2}\right| +12 \left(2 \tilde{q}^5-3 \mu  \tilde{q}^4-4 \mu ^3 \tilde{q}^2\right) \ln \left| \frac{\tilde{q}}{\tilde{q}-2 \mu }\right| \bigg] \nonumber\\
&&-\frac{e^2 \Theta(\tilde{q}-\mu)\Theta (2\mu-\tilde{q})}{288 \pi^2 \hbar^2 \tilde{q}^3 v_F\Omega}  \bigg[4 \Omega^2 (\mu -q) \left(58 \mu ^2+7 q^2-2 \mu  q+3 \Omega^2\right)  +12 \left(2 \tilde{q}^5-3 \mu  \tilde{q}^4-4 \mu ^3 \tilde{q}^2\right) \ln \left| \frac{\tilde{q}}{2 \mu -\tilde{q}}\right| \nonumber\\
&&  -6 \left(\tilde{q}^2-\Omega^2\right) \left(4 \mu ^3-2 \tilde{q}^3+3 \mu \tilde{q}^2+3 \mu  \Omega^2\right) \ln \left|\frac{(\tilde{q}-2 \mu )^2-\Omega^2}{\tilde{q}^2-\Omega^2}\right| \nonumber\\
&&+ 3 \Omega \left(\tilde{q}^2-\Omega^2\right) \left(12 \mu ^2+3 \tilde{q}^2+\Omega^2\right) \ln \left| \frac{(\Omega-\tilde{q}) (\Omega+2 \mu -\tilde{q})}{(\Omega+\tilde{q}) (\Omega-2 \mu +\tilde{q})}\right|\bigg],
\label{app-3-sigma-Im-xx-intra}
\end{eqnarray}

\begin{eqnarray}
&&\text{Im}\, \sigma_{\parallel}^{(5,p), \text{intra}}(\Omega,\mathbf{q}) = \frac{e^2 \Omega \Theta(\mu-\tilde{q})}{72 \pi^2  \hbar^2 \tilde{q}^3 v_F }\left[16 \tilde{q}^3-6 \tilde{q} \left(12 \mu ^2+\Omega^2\right) +3 \Omega \left(12 \mu ^2-3 \tilde{q}^2+\Omega^2\right) \ln \left| \frac{\Omega+\tilde{q}}{\Omega-\tilde{q}}\right| \right] \nonumber\\
&&+ \frac{e^2 \Omega \Theta(\tilde{q}-\mu)\Theta (2\mu-\tilde{q})}{72 \pi^2  \hbar^2 \tilde{q}^3 v_F }  \left[2 (2 \mu -\tilde{q}) \left(8 \tilde{q}^2+4 \mu \tilde{q}-3 \Omega^2-40 \mu ^2\right)+ 3 \Omega \left(12 \mu ^2-3 \tilde{q}^2+\Omega^2\right)\ln \left|\frac{\Omega+2 \mu -\tilde{q}}{\Omega-2 \mu +\tilde{q}}\right| \right] \nonumber\\
&& -\frac{e^2 \Omega \Theta(\tilde{q}-\mu)\Theta (2\mu-\tilde{q})}{144 \pi^2 \hbar^2 \tilde{q}^3 v_F}  \bigg[ 4 (\mu -\tilde{q} ) \left(8 \tilde{q}^2+2 \mu \tilde{q}-3 \Omega^2-58 \mu ^2\right)+3 \Omega \left(12 \mu ^2-3 \tilde{q}^2+\Omega^2\right) \ln \left|\frac{(\Omega-\tilde{q}) (\Omega+2 \mu -\tilde{q})}{(\Omega+\tilde{q}) (\Omega-2 \mu +\tilde{q})}\right| \nonumber\\
&&+6 \left(4 \mu ^3+\tilde{q}^3-3 \mu \tilde{q}^2+3 \mu \Omega^2\right) \ln \left|\frac{\tilde{q}^2-\Omega^2}{(\tilde{q}-2 \mu )^2-\Omega^2}\right| \bigg] -\frac{e^2 \Omega \Theta (\tilde{q}-2\mu)}{144 \pi^2 \hbar^2 \tilde{q}^3 v_F} \bigg[ 4 \mu  \left(22 \mu ^2-6 \tilde{q}^2+12 \mu \tilde{q}+3 \Omega^2\right) \nonumber\\
&&+3 \Omega \left(12 \mu ^2-3 \tilde{q}^2+\Omega^2\right) \ln \left| \frac{(\Omega-\tilde{q}) (\Omega-2 \mu +\tilde{q})}{(\Omega+\tilde{q}) (\Omega+2 \mu -\tilde{q})}\right| +6 \left(4 \mu ^3+\tilde{q}^3-3 \mu \tilde{q}^2+3 \mu \Omega^2\right) \ln \left|\frac{\tilde{q}^2-\Omega^2}{(\tilde{q}-2 \mu )^2-\Omega^2}\right| \bigg].
\label{app-3-sigma-Im-zz-intra}
\end{eqnarray}
Physically, the cutoff $\Lambda$ corresponds to the energy scale at which the relativistic-like dispersion relation of quasiparticles becomes inapplicable. In addition, $\sigma_{\perp}^{\rm (5,p)}(\Omega,\mathbf{q}) \equiv \sigma_{xx}^{(5,p)}(\Omega,\mathbf{q})$ and $\sigma_{\parallel}^{\rm (5,p)}(\Omega,\mathbf{q}) \equiv \sigma_{zz}^{(5,p)}(\Omega,\mathbf{q})$.

Since Eqs.~(\ref{app-3-sigma-Im-xx-inter})--(\ref{app-3-sigma-Im-zz-intra}) are cumbersome, let us discuss a few simplified cases. For $\mu=0$, we have
\begin{eqnarray}
\label{app-chiral-sigma-Im-inter-mu=0}
\text{Im}\, \sigma_{\perp}^{(5,p), \text{inter}}(\Omega,\mathbf{q})&=& - \frac{e^2 }{12 \pi^2  \hbar^2 v_F \Omega } \left[\left(\Omega ^2-\tilde{q}^2\right) \ln \left|\frac{\Lambda ^2-\Omega ^2}{\Omega ^2-\tilde{q}^2}\right|+2 \tilde{q}^2 \ln \left|\frac{\Lambda }{\tilde{q}}\right|\right], \\
\label{app-chiral-sigma-Im-intra-mu=0}
\text{Im}\, \sigma_{\parallel}^{(5,p), \text{inter}}(\Omega,\mathbf{q})&=&- \frac{e^2 \Omega}{12 \pi^2  \hbar^2 v_F  } \ln \left| \frac{\Lambda ^2-\Omega ^2}{\Omega ^2-\tilde{q}^2}\right|,\\
\text{Im}\, \sigma_{\perp}^{(5,p), \text{intra}}(\Omega,\mathbf{q})&=& \text{Im}\, \sigma_{\parallel}^{(5,p), \text{intra}}(\Omega,\mathbf{q}) =0.
\end{eqnarray}

Finally, let us show that our results agree with those in the literature in the corresponding limits. At $\tilde{q} \rightarrow 0$, we obtain
\begin{eqnarray}
\label{app-chiral-sigma-Im-inter-q=0}
\text{Im}\, \sigma_{\perp}^{(5,p), \text{inter}}(\Omega,\mathbf{0}) &=& \text{Im}\, \sigma_{\parallel}^{(5,p), \text{inter}}(\Omega,\mathbf{0})=
- \frac{e^2 \Omega }{12 \pi^2 \hbar^2  v_F}  \ln \left|\frac{\Lambda ^2-\Omega^2}{\Omega^2-4 \mu ^2}\right|,\\
\label{app-chiral-sigma-Im-intra-q=0}
\text{Im}\, \sigma_{\perp}^{(5,p), \text{intra}}(\Omega,\mathbf{0}) &=&\text{Im}\, \sigma_{\parallel}^{(5,p), \text{intra}}(\Omega,\mathbf{0}) =\frac{e^2 \mu ^2}{3 \pi^2 \hbar^2  v_F \Omega}.
\end{eqnarray}
These expressions agree with those derived in Refs.~\cite{Rosenstein:2013,Roy-DasSarma:2016} at $q=0$ for $\Lambda\gg \Omega$.

\bibliography{library_short}

\end{document}